\documentclass[referee]{aa}    
\usepackage{txfonts}
\usepackage{aalongtable}
\usepackage{graphicx}

\begin{document}

\title{X--ray spectral properties of AGN in the Chandra Deep Field
South}

\author{P. Tozzi\inst{1} \and R. Gilli\inst{2}\and V. Mainieri\inst{3}
\and C. Norman\inst{4,5} \and G. Risaliti\inst{2,6} \and
P. Rosati\inst{7} \and J. Bergeron\inst{8} \and S. Borgani\inst{9}
\and R. Giacconi\inst{4} \and G. Hasinger\inst{3} \and
M. Nonino\inst{1} \and A. Streblyanska\inst{3} \and
G. Szokoly\inst{3} \and J.X. Wang\inst{10} \and W. Zheng\inst{4}}

\institute{INAF Osservatorio Astronomico di Trieste, via G.B. Tiepolo
11, I--34131, Trieste, Italy \and INAF Osservatorio Astrofisico di
Arcetri, Largo E. Fermi 5, 50125 Firenze, Italy \and
Max--Planck--Institut f\"ur extraterrestrische Physik,
Giessenbachstra\ss e, PF 1312, 85741 Garching, Germany \and Dept. of
Physics and Astronomy, The Johns Hopkins University, Baltimore, MD
21218, USA \and Space Telescope Science Institute, 3700 S. Martin
Drive, Baltimore, MD 21210, USA \and Harvard-Smithsonian Center for
Astrophysics, 60 Garden Street Cambridge, MA 02138 \and European
Southern Observatory, Karl-Schwarzschild-Strasse 2, D-85748 Garching,
Germany \and Institut d'Astrophysique de Paris, 98bis, bd Arago -
75014 Paris France \and Universit\`a di Trieste, Dip. Astronomia, via
Tiepolo 11, I--34131, Trieste, Italy \and Center for Astrophysics,
University of Science and Technology of China, Hefei, Anhui, 230026,
P. R. China}

\date{Received  / Accepted}

\abstract{We present a detailed X--ray spectral analysis of the
sources in the 1Ms catalog of the Chandra Deep Field South (CDFS)
taking advantage of optical spectroscopy and photometric redshifts for
321 extragalactic sources out of the total sample of 347 sources.  As
a default spectral model, we adopt a power law with slope $\Gamma$
with an intrinsic redshifted absorption $N_H$, a fixed Galactic
absorption and an unresolved Fe emission line.  For 82 X--ray bright
sources, we are able to perform the X--ray spectral analysis leaving
both $\Gamma$ and $N_H$ free.  The weighted mean value for the slope
of the power law is $\langle \Gamma \rangle \simeq 1.75 \pm 0.02$, and
the distribution of best fit values shows an intrinsic dispersion of
$\sigma_{int} \simeq 0.30$.  We do not find hints of a correlation
between the spectral index $\Gamma$ and the intrinsic absorption
column density $N_H$.

We then investigate the absorption distribution for the whole sample,
deriving the $N_H$ values in faint sources by fixing $\Gamma = 1.8$.
We also allow for the presence of a scattered component at soft
energies with the same slope of the main power law, and for a pure
reflection spectrum typical of Compton--thick AGN.  We detect the
presence of a scattered soft component in 8 sources; we also identify
14 sources showing a reflection--dominated spectrum.  The latter are
referred to as Compton--thick AGN candidates.

By correcting for both incompleteness and sampling--volume effects, we
recover the intrinsic $N_H$ distribution representative of the whole
AGN population, $f(N_H) dN_H$, from the observed one.  $f(N_H)$ shows
a lognormal shape, peaking around $log(N_H)\simeq 23.1$ and with
$\sigma \simeq 1.1$.  Interestingly, such a distribution shows
continuity between the population of Compton--thin and that of
Compton--thick AGN.

We find that the fraction of absorbed sources (with $N_H>10^{22}$
cm$^{-2}$) in the sample is constant (at the level of about $75$\%) or
moderately increasing with redshift.
Finally, we compare the optical classification to the X--ray
spectral properties, confirming that the correspondence of unabsorbed
(absorbed) X--ray sources to optical Type I (Type II) AGN is accurate
for at least 80\% of the sources with spectral identification (1/3 of
the total X-ray sample).  

 }

\titlerunning{X--ray spectral properties  of AGN in the CDFS}
\maketitle

\section{Introduction}

Deep X--ray surveys with Chandra (Brandt et al. 2001; Rosati et
al. 2002; Cowie et al. 2002; Alexander et al. 2003; Barger et
al. 2003) and XMM (Hasinger et al. 2001) showed that the so called
X--ray background (XRB) is mainly provided by Active Galactic Nuclei
(AGN) both in the soft (0.5--2 keV) and in the hard (2--10 keV)
band. In particular, major progress has been made in the hard band,
for which the sources known before Chandra were providing only $\sim
30$\% of the XRB (Cagnoni et al. 1998; Ueda et al. 1999a).  While some
evidence of spectral hardening was found towards faint fluxes
(e.g. Della Ceca et al. 1999), most of the X-ray sources were
identified with Broad Line AGN with a typical X-ray spectral slope of
$\Gamma =1.7-2.0$, steeper than that of the XRB ($\Gamma \simeq 1.4$).
On the contrary, the source population discovered by Chandra and XMM
at fluxes below $\sim 10^{-13}-10^{-14}$ erg cm$^{-2}$ s$^{-1}$ in the
hard band, is constituted mostly by obscured AGN with a hard spectrum,
and provides the solution to the ``spectral paradox'' as predicted by
the XRB synthesis models (Setti \& Woltjer 1989, Madau, Ghisellini \&
Fabian 1993; Comastri, Setti, Zamorani \& Hasinger 1995; Gilli,
Salvati \& Hasinger 2001).  The detection limit reached in the hard
band in the 2Ms exposure of the Chandra Deep Field North is $S \simeq
2 \times 10^{-16}$ erg s$^{-1}$ cm$^{-2}$ (Alexander et al. 2003) and
a factor 2 higher in the 1Ms exposure of the Chandra Deep Field South
(CDFS, Rosati et al. 2002; Giacconi et al. 2002).  The XRB is now
resolved at the level of $\sim 80$\% in the 1--2 and 2--8 keV bands
(see Hickox \& Markevitch 2005), with the AGN providing the large
majority of the resolved fraction.  While a non--negligible part of
the unresolved fraction in the soft band is expected to be contributed
by a diffuse warm intergalactic medium (e.g., Cen \& Ostriker 1999),
Worsley et al. (2004; 2005) pointed out that at $E> 6$ keV less and
less of the hard XRB is resolved, showing that a significant
population of strongly absorbed, possibly Compton--thick sources,
preferentially at $z<1$, is still not observed (see also Comastri
2004; Brandt \& Hasinger 2005).

The two Chandra Deep Field Surveys lead to the detection of several
populations of X--ray extragalactic sources: unabsorbed AGN (defined
as sources with absorbing column densities $N_H < 10^{22}$ cm$^{-2}$),
usually identified with optical Broad Line (Type I) AGN and QSO;
absorbed AGN (with column densities $N_H \geq 10^{22}$ cm$^{-2}$),
optically identified mostly as narrow line (Type II) AGN, distributed
around moderate redshifts $z\sim 1$ (see Barger et al. 2002; Szokoly
et al. 2004); X--ray bright, optically normal galaxies (XBONG, see
Comastri et al. 2001) which generally harbor obscured AGNs; high
redshift Type II QSO (see Norman et al. 2002; Stern et al. 2002;
Mainieri et al. 2005a; Ptak et al. 2005); starburst and quiescent
galaxies at $z<1$ (Bauer et al. 2002; Hornschemeier et al. 2003;
Norman et al. 2004), which contribute to the XRB only $2-3$\% in
energy, but they are expected to outnumber the AGN at fluxes below
$1\times 10^{-17}$ erg cm$^{-2}$ s$^{-1}$ (Bauer et al. 2004a).  In
this Paper we will focus on the X--ray properties of the AGN
population, in order to provide a baseline for possible models of the
AGN formation and evolution.

Tentatively, the different classes of AGN--powered X--ray sources can
be associated to three phases: a first phase of strong accretion onto
the massive black hole, characterized by high intrinsic absorption and
intense star formation (for recent evidence in the submm range see
Alexander et al. 2005a), followed by an unobscured phase, and
subsequent fading (see Fabian 1999; Granato et al. 2004).  A test of
this or other possible scenarios for the accretion history and galaxy
formation in the Universe, requires a good knowledge of the
distribution of the X--ray properties of the AGN population, in
particular intrinsic luminosity and intrinsic absorption as a function
of cosmic epoch, as well as their relation with the optical
properties.  The distribution of the intrinsic absorption, $N_H$, is
known only for local, optically selected Seyfert II galaxies (Risaliti
et al. 1999).  These local samples, selected to be complete as a
function of intrinsic luminosity, typically include medium or low
luminosity sources, and about 50\% of them are Compton--thick.
Difficulty of assembling large unbiased AGN sample as a function of
intrinsic luminosity, has hampered attempts to measure the $N_H$
distribution.  The $N_H$ distribution and the evolution of the
fraction of absorbed sources, has been investigated recently by Ueda
et al. (2003) from a combination of surveys from HEAO1, ASCA and
Chandra.  Their sample is dominated by bright, low absorption AGN, and
their $N_H$ distribution is broadly peaked above $N_H > 10^{22}$
cm$^{-2}$.  Except for a few objects with good photon statistics, Ueda
et al. use the redshift and the hardness ratio to derive the intrinsic
luminosity distribution in the 2--10 keV band as a function of
redshift, without performing a single--source analysis.  Similar
results have been recently obtained by La Franca et al. (2005) on the
basis of the HELLAS2XMM sample combined with other catalogs.  At
brighter fluxes, other investigations are under way both with Chandra
and XMM in wide, shallower surveys (ChaMP, Green et al. 2004;
Silverman et al. 2005; XMM--BSS, Della Ceca et al. 2004; CLASXS, Yang
et al. 2004, Steffen et al. 2004; HELLAS2XMM, Baldi et al. 2002;
Perola et al. 2004).  We believe that these X--ray surveys, designed
to bridge the gap between the pencil beam, deepest surveys and the
wide shallow ones from previous missions, are probably biased against
heavily absorbed faint AGN, whose fraction is expected to increase
towards fainter fluxes.  On the other hand, optical surveys can
actually discover heavily obscured AGNs at moderate redshift ($z<1.3$)
but only through extensive optical spectroscopy of large sample of
galaxies, such as the SDSS, among which type II AGNs can be identified
from the strong narrow emission lines (for example,
[OII]$\lambda$3727$\AA$ or [OIII]$\lambda$5007$\AA$).  In the absence
of high--sensitivity X--ray surveys above 10 keV, we propose that the
search for the still missing strongly absorbed AGN population can be
best performed through a detailed spectral analysis of faint sources
detected in very deep X--ray surveys.

In this Paper, we present a systematic study of the X--ray spectra of
all the sources in the CDFS, taking advantage of spectroscopic
(Szokoly et al. 2004) and photometric (Zheng et al. 2004; Mainieri et
al. 2005a) redshifts from the optical follow--up program with the
ESO--VLT.  Given the flux limits in the CDFS ($5.5 \times 10^{-17}$
and $4.5\times 10^{-16}$ erg cm$^{-2}$ s$^{-1}$ in the soft and hard
band respectively), the 347 sources detected (346 from the catalog of
Giacconi et al. 2002 plus one added in Szokoly et al. 2004) are mostly
AGN, with a fewer number of normal or star forming galaxies with
respect to CDFN, where, thanks to the lower flux limits, normal
galaxies start to be a significant fraction of the faint source
population.  The Paper is structured as follows.  In \S2 we briefly
describe the X--ray and the Optical data.  In \S3 we describe our
X--ray spectral analysis procedure, after dividing the sample into two
subsamples based on the counts statistics.  In \S4 we present the
X--ray spectral analysis of the X--ray bright sample, focusing on the
slope of the power law component.  In \S5 we present the X--ray
spectral analysis for the whole sample of 321 sources with measured
redshift and total luminosity $L_X> 10^{41}$ erg s$^{-1}$ (we exclude
the faintest luminosity bin which is doninated by normal galaxies),
focusing on the intrinsic absorption.  In \S6 we discuss the
distributions of the X--ray spectral properties after correcting for
incompleteness and sampling--volume effects, deriving in particular
the intrinsic absorption distribution.  This allows us to estimate the
fraction of absorbed sources in our sample as a function of epoch.
Finally, in \S7 we compare the X--ray and optical properties,
revisiting the comparison of the Optical vs X--ray classification
scheme proposed by Szokoly et al. (2004).  Our conclusions are
summarized in \S8.  Luminosities are quoted for a flat cosmology with
$\Lambda=0.7$ and $H_0=70$ km/s/Mpc (see Spergel et al. 2003).

\section{The data}

The 1Ms dataset of the CDFS is the result of the coaddition of 11
individual {\it Chandra} ACIS--I (Garmire et al. 1992; Bautz et
al. 1998) exposures with aimpoints only a few arcsec from each other.
The nominal aim point of the CDFS is $\alpha=$3:32:28.0,
$\delta=-$27:48:30 (J2000).  The reduction and analysis of the X--ray
data are described in Giacconi et al. (2001), Tozzi et al. (2001) and
Rosati et al. (2002).  The final image covers 0.108 deg$^2$, where 347
X--ray sources are identified (the catalog is presented in Giacconi et
al. 2002).  Here we use an updated X-ray data reduction, where we used
Ciao 3.0.1 and CALDB2.26, therefore including the correction for the
degraded effective area of ACIS--I chips due to material accumulated
on the ACIS optical blocking filter at the epoch of the observation.
We also apply the recently released, time--dependent gain
correction\footnote{see
http://asc.harvard.edu/ciao/threads/acistimegain/}.

We briefly recall the main steps of the spectral analysis of the
reduced data.  First we extract the photon files and the spectrum
({\tt pha} file) for every source in our catalog, along with the
corresponding background.  The area of extraction of each source, as
described in Giacconi et al. (2001), is defined as a circle of radius
$R_s=2.4\times FWHM$ (with a minimum radius of 5 arcsec).  The FWHM is
modeled as a function of the off--axis angle to reproduce the
broadening of the PSF.  The background is extracted from an annulus
with outer radius $R_S+12''$ and an inner radius of $R_S+2''$, after
masking out other sources.  Each background spectrum samples more than
400 photons in the 0.5--7 keV range.  We create a response matrix and
an ancillary response matrix for each source.  To do that, we first
create the two matrices in the source position in each of the 11
observations of the CDFS (therefore the effect of the degraded
effective area of ACIS--I chips is applied individually to each
pointing).  Finally we sum the 11 files weighting them for the
exposure time of each exposure.  We notice that most of the sources
show variability (see Paolillo et al. 2004), therefore our measured
fluxes and luminosities are time--averaged on the observation
epochs. We also stress that, assuming there is no significant changes
in the spectra, we correctly measure the spectral shape of each
source, since the response matrices are time--averaged on the same
epochs, keeping track in the most detailed way of the characteristics
of the different regions and the different conditions of the detector
at the time of the observations.

The spectroscopic identification program carried out with the ESO-VLT
is presented in Szokoly et al. (2004).  The optical classification is
based on the detection of high ionization emission lines.  The
presence of broad emission lines (FWHM larger than 2000 km/s) like
Mg$_{II}$, C$_{III}$, and, at large redshifts, C$_{IV}$ and
Ly$\alpha$, identifies the source as a Broad Line AGN (BLAGN), Type--1
AGN or QSO in the simple unification model (Antonucci 1993). The
presence of unresolved high ionization emission lines (like O$_{III}$,
Ne$_V$, Ne$_{III}$ or He$_{II}$) identifies the source as a High
Excitation line galaxy (HEX), often implying an optical Type--2
classification.  Objects with unresolved emission lines consistent
with an H$_{II}$ region spectrum are classified as Low Excitation Line
galaxies (LEX), implying sources without signs of nuclear activity in
the optical (however, discriminating between a Seyfert II galaxy and
an $H II$ region galaxy involves the measure of line ratio as shown in
Veilleux \& Osterbrock (1987), which is not used here as a
classification scheme, considering also that their classification
scheme relies on lines which are not visible in optical spectra from
the ground at $z>0.7$).  Objects with typical galaxy spectrum showing
only absorption lines are classified as ABS; among the last two
classes we expect to find star--forming galaxies or Narrow Line
Emission Galaxies, but also hidden AGN.  The optical identification is
flagged according to the quality of the optical information.  Quality
flags $Q\geq 1$ indicates spectroscopic redshifts (see Table
\ref{results}).  In several cases, the optical spectral properties do
not allow us to obtain a secure determination of the spectral type.
As shown in Szokoly et al. (2004), the optical classification scheme
is failing in identifying an AGN in about 40\% of the X-ray sources
optically classified as LEX or ABS.  Therefore, an X--ray
classification scheme, based on the source hardness ratio and observed
X--ray luminosity, was worked out by Szokoly et al. (2004) and
compared with the optical one (see their Fig.~13).  In Section \S 7 we
will reconsider this X--ray classification scheme using the intrinsic
luminosities (as opposed to observed ones) and intrinsic absorption
(as opposed to the hardness ratio).

Optical and near-IR images of the CDFS are also used to derive
photometric redshifts for all the remaining X--ray sources.  Using the
widest multiwavelength photometry available today, Zheng et al. (2004)
and Mainieri et al. (2005a) derived photo--z for the whole sample of
sources but four.  Photometric redshifts are obtained from different
methods labelled with different quality flags (see Zheng et al. 2004
for details).  When we have consistent redshift from more than one
method, the corresponding quality flag is the sum of the single $Q$
(always less than 1 for photometric redshift). Given the good
agreement of photometric redshifts with spectroscopic ones (see Zheng
et al. 2004), we do not divide our sample according to the optical
spectra quality.  Indeed, our statistical analysis is not expected to
be significantly affected by uncertainties in the photometric
redshifts.  Uncertainties in the redshift estimate may instead
significantly affect the search for the Fe line, as we discuss later.

The total number of sources with spectral or photometric redshift
$z>0$ is 336 over a total of 347 X--ray detections.  Besides the 4
X-ray sources without any redshift estimate, we indeed identify 7
stars with good optical spectra.  Therefore the spectral completeness
of our sample of extragalactic sources is $\sim 99$ \%.  Since we want
to focus on AGN, we adopt a conservative criterion and exclude 15
sources with total luminosity in the 0.5--10 keV band $L_X< 10^{41}$
erg s$^{-1}$, a luminosity range which is expected to be dominated by
normal or star forming galaxies.   We note that the higher
luminosiy range $10^{41} < L_X < 10^{42}$ erg s$^{-1}$ may include
several star forming galaxies as well, with star formation rate of the
order of $100 M_\odot$/yr.  However, we keep all the sources in the
luminosiy range $10^{41} < L_X < 10^{42}$ erg s$^{-1}$ to include any
possible low--luminosity AGN in the sample.  The final sample amounts
to 321 sources.  The redshifts with the corresponding spectral quality
are shown along with the results from the X--ray spectral fits in
Table \ref{results}.

\section{The X--ray spectral analysis}

\subsection{Fitting strategy}

\begin{figure}
\resizebox{\hsize}{!}{\includegraphics{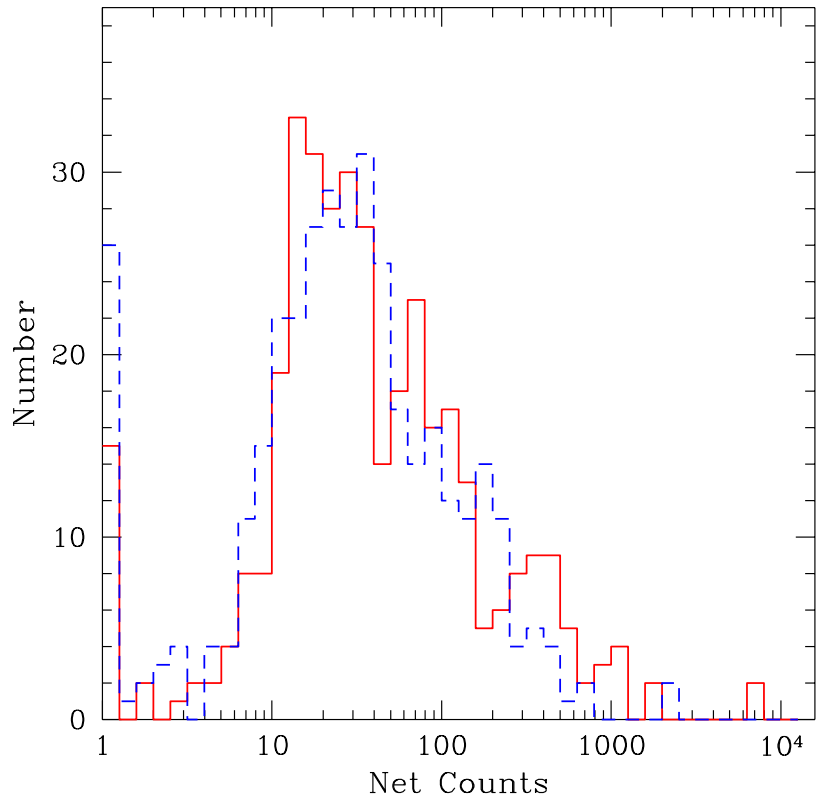}}
\caption{Distribution of the net detected counts for all the sources
in the sample (solid line: 0.5--2 keV band counts; dashed line: 2--7
keV band counts).}
\label{counts_distrib}
\end{figure}

We use XSPEC v11.3.1 (see Arnaud 1996) to perform the spectral fits.
The ability of obtaining a reliable fit depends on the X--ray spectral
quality, or, in simpler terms, on the signal to noise of the spectrum
under analysis.  The distribution of the net counts in the 0.5--7 keV band
for all the sources in our sample, peaks below $\simeq 100$ (see
Figure \ref{counts_distrib}).  The mean value of the net detected
counts in the total 0.5--7 keV band for all the sources in our sample
(including the two X--ray brightest sources in the sample, with about
10000 counts each) is $\simeq 240$ counts, while the median is much
lower $\simeq 70$ counts.

\begin{figure}
\resizebox{\hsize}{!}{\includegraphics{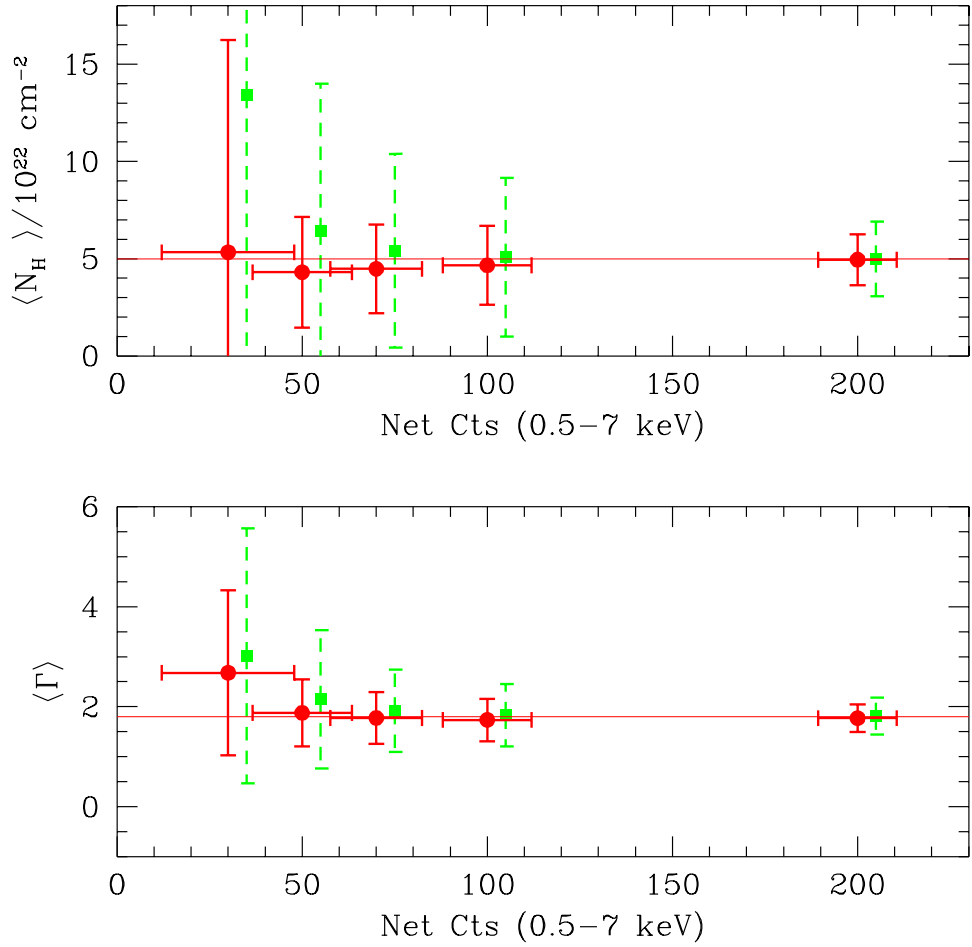}}
\caption{The average best--fit parameters (with rms dispersion) for a
source with intrinsic $N_H = 5\times 10^{22}$ cm$^{-2}$ and $\Gamma =
1.7$ at $z=1$ fitted with Cash statistics (filled circles, continuous
error bars) and $\chi^2$ (filled squares, dashed error bars) versus
the number of net detected counts in the 0.5-7 keV band.  The $\chi^2$
points are slightly shifted along the x--axis for clarity.  The same
source is simulated 1000 times for five different intrinsic
normalizations, resulting in a different average number of net
detected counts.
\label{compare}}
\end{figure}

Therefore, the strategy for the X--ray spectral analysis must be
appropriate for the low counts regime.  In performing the spectral
fits we used an extension of the Cash statistics which makes use of
both the source and background spectral files\footnote{see {\tt
http://heasarc.gsfc.nasa.gov/docs/xanadu/xspec/manual/node57.html}}.
Cash statistics is applied to unbinned data, and therefore exploit the
full spectral resolution of the ACIS--I instrument, allowing better
performance with respect to the canonical $\chi^2$ analysis,
particularly for low signal--to--noise spectra (Nousek \& Shue
1989). In order to assess the ability of our fitting procedure in a
typical case (a source with $\Gamma = 1.7$ and $N_H =5 \times 10^{22}$
cm$^{-2}$ at $z=1$) we run several simulations for different input
fluxes, in which we try to recover the input parameters with two
different fitting procedures: Cash statistics (unbinned) and the
classic $\chi^2$ statistics with a binning of 10 photons per bin. The
results are summarized in Figure \ref{compare}.  Note that we are
forced to use a binning of 10 photons (as opposed to the commonly used
binning of 20 photons) in order have a reasonable number of bins to
perform the $\chi^2$ fits in the low--counts regime.  Such a small
binning is known to give inappropriate weights for the $\chi^2$
analysis, therefore we do not mean to present a detailed comparison of
the two methods.  Indeed, here we just explore the effects that their
use would have in the spectral analysis of our sample.  For the
$\chi^2$ statistics, we find that for sources with a number of net
counts equal or larger than 50, the input parameters are recovered
with very good accuracy, while for lower values, the peak of the
distribution of the best--fit--values starts to depart from the input
value.  The shift in the distribution of the best--fit values is a
consequence of the binning, which, especially in the case of
low--counts statistics, acts as an effective smoothing on the
spectrum.  On the other hand, the distribution of the best--fit values
with Cash--statistics appears to be closer to the input values.  In
addition, the {\sl rms} dispersion of best--fit values is
significantly lower with respect to the $\chi^2$ statistics. We also
checked that the confidence levels for the Cash--statistics can be
defined as in the $\chi^2$--statistics (i.e., $\Delta C =1.0$
corresponds to 1 $\sigma$, $\Delta C =2.7$ corresponds to 90\%
c.l. for one interesting parameter).  Therefore we choose to quote
only the best fit values obtained with the Cash statistics.

\begin{figure}
\resizebox{\hsize}{!}{\includegraphics{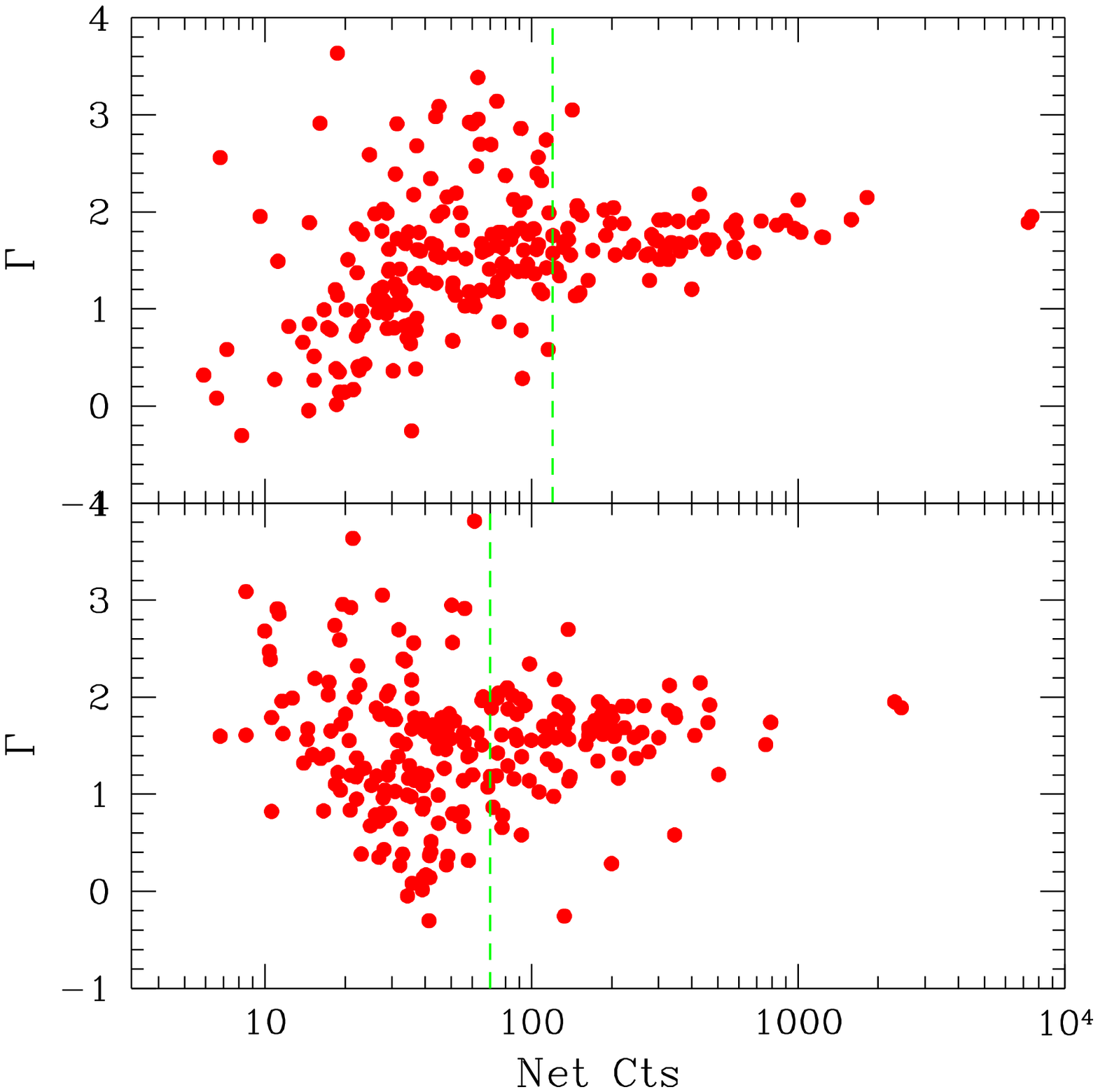}\includegraphics{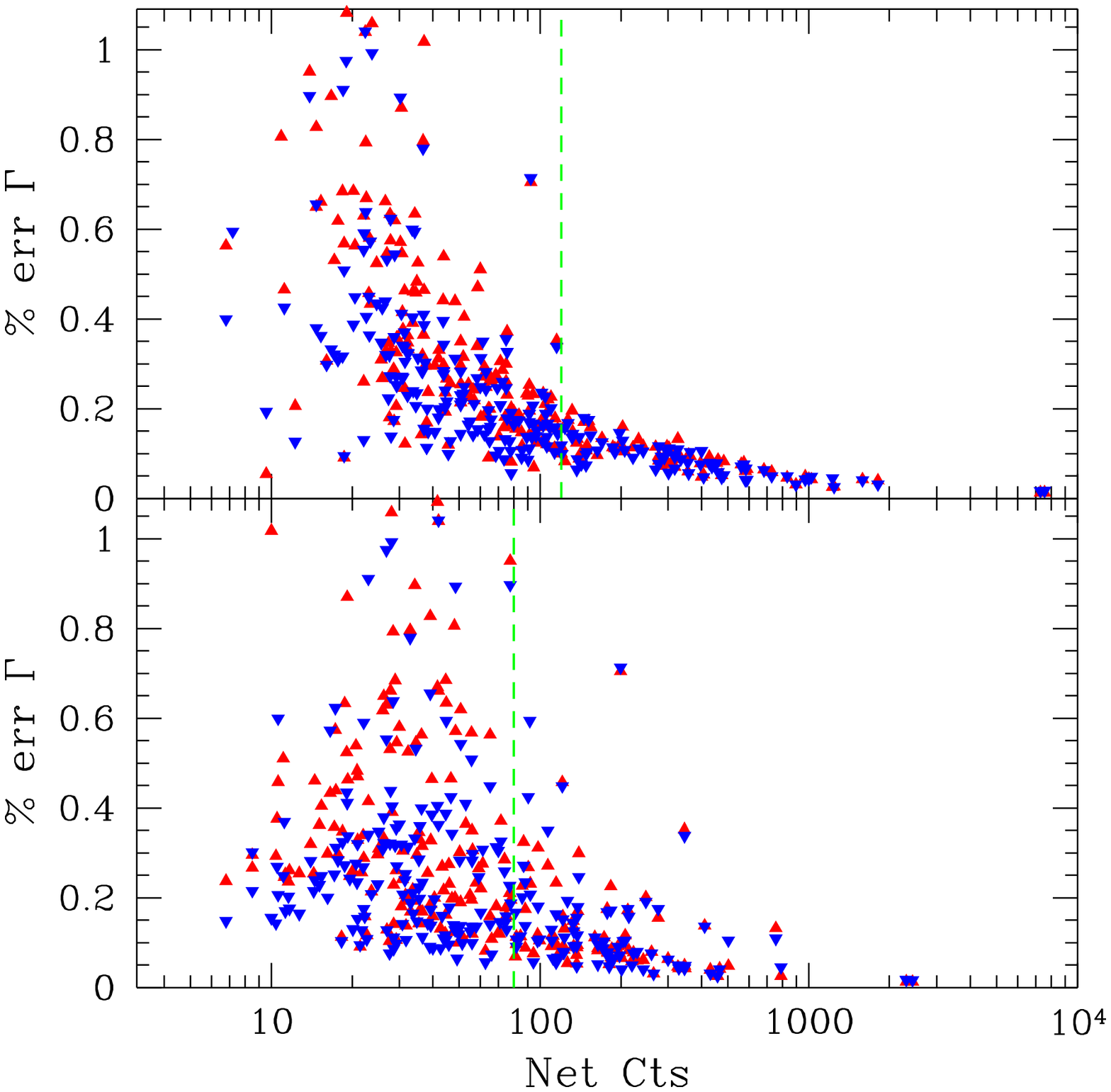}}
\caption{Left: best fit values of the spectral slope $\Gamma$ as a
function of the net counts in the soft (top panel) and hard (bottom
panel) bands (for sources with more than 40 net counts in the 0.5--7
keV band).  Right: fractional statistical error ($1 \sigma$ c.l.) on
$\Gamma$ as a function of net counts in the soft (top panel) and hard
(bottom panel) band (upper and lower errors are shown as triangles and
upside--down triangles respectively).  Vertical dashed lines are the
thresholds adopted to select the bright X--ray sample (82 sources with
more than 120 net counts in the soft band or more than 80 in the hard
band, or more than 170 counts overall).
\label{cts_g}}
\end{figure}

Of course, the weak signal of our faintest sources limits the ability
to perform a fit keeping all the spectral parameters free.  To
determine the validity of our approach, we first run the fit for our
default model with three free parameters ($N_H$, $\Gamma$ and
normalization) on all the sources with more than 40 net detected
counts in the total 0.5--7 keV band\footnote{Given the low background
of Chandra and the small extraction regions used for the sources, the
correlation between signal--to--noise in a given band and total net
counts is very tight.  Therefore for simplicity we select our sources
on the basis of the net detected counts.  }.  First we focus on the
distribution of the best--fit values for $\Gamma$ as a function of the
net counts (see Figure \ref{cts_g}, left).  We notice that for sources
detected with a large number of counts (larger than $\simeq 200$) the
spectral slope is almost constant.  On the other hand, at low counts,
the best fit spectral slope $\Gamma$ shows an apparent trend
associated with a significant increase in the dispersion on $\Gamma$
(see Figure \ref{cts_g}, right): lower values at lower soft counts,
higher values at lower hard counts.  In principle this is expected,
since most of the sources with few soft counts are among the hardest
sources, and they can be fitted with a flat power law, and viceversa
the softest sources can be fitted with a very steep power
law. However, we argue that this behaviour may be affected by the poor
statistics.  To avoid any possible bias induced by the low statistics,
we conservatively define an X-ray bright sample by considering those
sources exceeding at least one of these thesholds: 170 total counts,
120 soft counts, 80 hard counts.  As we can see in Figure
\ref{cts_g}b, the threshold on the soft counts is particularly
efficient in selecting sources for which the statistical error on
$\Gamma$ is smaller than 20\% (about 10\% in average).  The bright
sample, constituted by 82 sources, will be used to investigate both
the intrinsic spectral slope $\Gamma$ and the intrinsic absorption
$N_H$.  We remark here that the bright sources are selected on the
basis of the net detected counts, and not on the basis of the energy
flux; among the brigth sample, we find sources with fluxes larger than
$4\times 10^{-16}$ erg s$^{-1}$ cm$^{-2}$ in the soft and $1.3 \times
10^{-15}$ erg s$^{-1}$ cm$^{-2}$ in the hard band.  As for the
remaining 3/4 of the sample, we decide to fix the slope to the
canonical value of $\Gamma \simeq 1.8$ (see Turner et al. 1997), which
is, in turn, very close to the average value measured for our bright
sample (as shown in \S 4), and focus on the intrinsic absorption.

\subsection{Spectral models}

We assume a default spectral model based on a power law (XSPEC model
{\tt pow}) and intrinsic absorption at the source redshift (XSPEC
model {\tt zwabs}) with redshift frozen to the spectroscopic or
photometric value.  Also, we search for the Fe K$\alpha$ line at 6.4
keV rest--frame, which is one of the most common features of AGN
X--ray spectra.  To investigate the presence of such a line, we added
a redshifted unresolved Gaussian line at $6.4/(1+z)$ keV (Nandra \&
Pounds 1994).  We also take into account the local Galactic absorption
(XSPEC model {\tt tbabs}) with a column density frozen to $N_H = 8
\times 10^{19}$ cm$^{-2}$ (from Dickey \& Lockman 1990).  The fits are
performed on the energy range 0.6--7 keV.  We cut below 0.6 keV to
avoid uncertainties in the ACIS calibration in an energy range which
anyway offers a small effective area.  At high energies, the
efficiency of Chandra is rapidly decreasing, and the energy bins at
more than 7 keV are dominated by the noise for the large majority of
the sources in our flux range.  It has recently been shown that a
methylen layer on the Chandra mirrors increases the effective area at
energies larger than 2 keV (see Marshall et al. 2003)\footnote{ see
http://cxc.harvard.edu/ccw/proceedings/03\_proc/presentations/marshall2}.
This has a small effect on the total measured fluxes, but it can have
a non-negligible effect on the spectral parameters.  To correct for
this, we include in the fitting model a ``positive absorption edge''
(XSPEC model {\tt edge}) at an energy of 2.07 keV and with $\tau =
-0.17$ (Vikhlinin et al. 2005).  This multiplicative component
artificially increases the hard fluxes by $\simeq 3.5$\%, therefore
the final hard fluxes and luminosities computed from the fit are
corrected downwards by the same amount.

In some cases, the fit with a simple absorbed power law may not be a
good description of the X--ray spectrum.  On the other hand, our
limited counts statistics does not allow us to investigate for complex
spectral shapes as often observed in AGN.  However, we identify two
possible additional spectral models.  A first spectral model we
investigate is the presence of a soft component in addition to the
absorbed power law, as often found in the X-ray spectra of Seyfert 2
galaxies (e.g. Turner et al. 1997).  Such a soft component can arise
from several physical processes, like nuclear radiation scattered by a
warm medium (the so-called "warm mirror", e.g.  Matt et al. 1996), or
nuclear radiation leaking through the absorber. In this cases, the
soft component is expected to have the same spectral slope of the main
power law.  Here we do not consider the soft excess possibly due to
thermal emission or comptonization of soft photons, as found in bright
quasars (see Porquet et al. 2004).  Thus, we repeated the fits simply
adding to the Compton--thin model an unabsorbed power law component
with slope equal to that of the main power law, requiring the
intrinsic normalization of the soft component to be always less than
10\% of the intrinsic normalization of the main power law.  This last
requirement embraces typical values both for a scattered component and
for leaky absorbers (see Turner et al. 1997).  This upper limit may
exclude some leaky absorber with a low covering fraction, but at the
same time helps us in avoiding false detections of high--normalization
soft components implying spuriously high values of $N_H$ relative to
the absorbed component.  With this procedure, a soft component is
detected with $\Delta C > 2.7$ in 8 sources.

Moreover, when the intrinsic absorption is as high as $N_H\simeq
1.5\times 10^{24}$ cm$^{-2}$, the Compton optical depth is equal to
unity and the directly transmitted nuclear emission is strongly
suppressed in the Chandra soft and hard bands.  In particular, for an
intrinsic power--law spectrum with $\Gamma = 1.8$, the fraction of
transmitted photons is less than $2\%$ in the soft band up to redshift
$z=2$.  Absorption is less severe in the hard band, where for $z>1$
already a fraction of $10\%$ of the emitted photons are recovered.  It
is clear that only the intrinsically brightest, heavily absorbed
high--redshift AGN can be detected by their transmitted nuclear
emission.  In this regime, a radiation component reflected by a cold
medium, expected to be in average $6\%$ of the intrinsic power in the
2--10 keV band, starts to be important.  For these Compton--thick
sources, the most commonly observed spectrum is dominated by a
Compton--reflection continuum from cold medium, usually assumed to be
produced by the far inner side of the putative obscuring torus.  This
can be modeled with the XSPEC model {\tt pexrav} (Magdziarz \&
Zdziarski 1995) plus the redshifted Fe K line.

The {\tt pexrav} model often provides a better fit for the sources in
our sample with a flat spectrum.  For simplicity, we fix all the
parameters to the default, typical values ($\Gamma=1.8$, reflection
relative normalization=0, element and Fe abundance set to 1, cosine of
inclination angle set to 0.45) but the normalization of the intrinsic
power law spectrum.  Our selection of Compton--thick candidates, then,
is based on the comparison of the Cash--statistics obtained in the
best fits with the {\tt zwabs pow} model (with two free parameters,
$N_H$ and normalization) with that obtained with the pure reflection
model (with only one free parameter, the normalization).  The
difference $\Delta C$ is an indication of the goodness of the {\tt
pexrav} model with respect to the standard absorbed power law.  Due to
the different number of free parameters and the low signal-to-noise
typical of our sources, we choose a threshold $\tilde \Delta C$ to
select Compton--thick candidates after extensive simulations.  The
simulations procedure is described in Appendix B.  We find that a
threshold $\tilde \Delta C=2$ allows us to select a sample of
Compton--thick candidates with a contamination fraction of about 20\%.
On the other hand, we also find that with our selection criteria, we
may miss a fraction as high as 40\% of the total Compton--thick
population.  Indeed, we find that, given the typical signal--to--noise
of our sample, it is extremely difficult to efficiently select
Compton--thick sources on the basis of the shape of the X--ray
spectrum.  We recognize that, in order to perform a careful search for
Compton--thick candidates, other spectral features, like the Fe K
line, or other wavelengths (like the submillimeter range of SCUBA)
should be explored (see Alexander et al. 2005b).  This goes beyond the
goal of this Paper.

To summarize, we label as C--thin the sources for which the best fit
model is a power law with intrinsic absorption; C--thick the sources
for which the best fit is given by a {\tt pexrav} model; finally
Soft--C for sources whose best fit model includes a soft component
with the same slope of the main power law.  Finally, we always add a
gaussian component to model the Fe K line, which, in case of no
detection, gives a null or negligible contribution to the spectral
shape.

\section{Spectral slope for the bright sample}

\begin{figure}
\resizebox{\hsize}{!}{\includegraphics{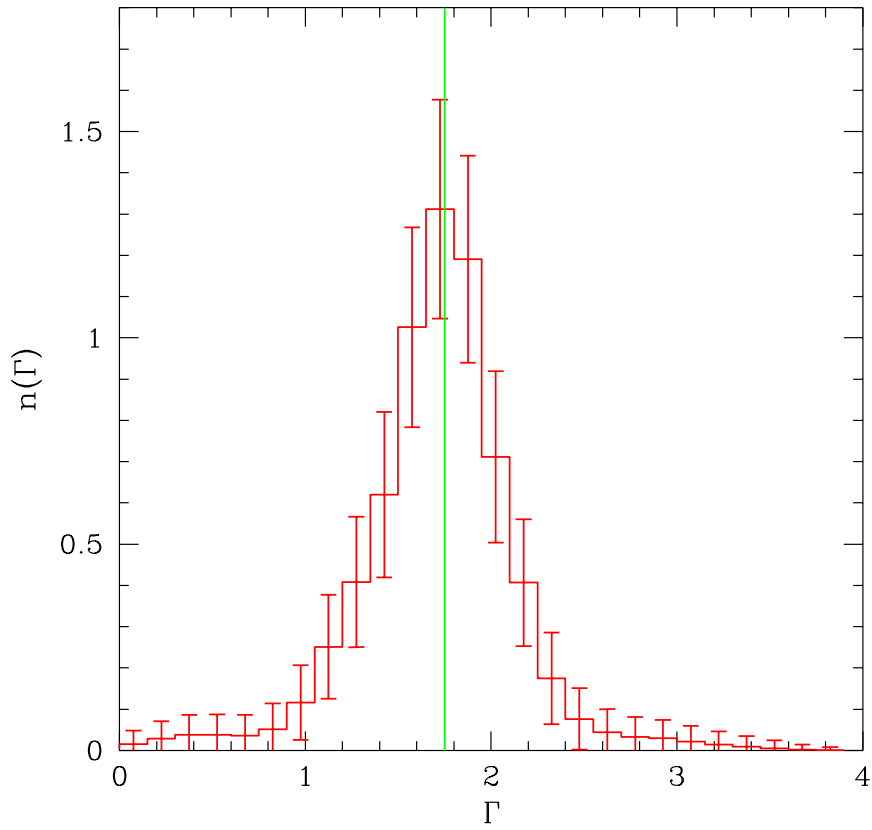}}
\caption{Distribution of the spectral slope $\Gamma$ for the X--ray
bright sample (82/321 sources).  Error bars are 1 $\sigma$ poissonian
errors.  The solid vertical line refers to the central value $\langle
\Gamma \rangle =1.75$.
\label{g_histo}}
\end{figure}

First, we consider only the X--ray bright sample of 82 sources with
more than 120 net detected counts in the soft band or more than 80 in
the hard band, and more than 170 net counts overall.  Among them, only
two sources with soft component are found, and no Compton thick
candidates.  We note that the low fraction of sources with significant
soft component, lower than that in the local sample of Turner et
al. (1997), may be ascribed to the high redshifts in our sample, for
which the soft component is often shifted below 0.6 keV.  We use this
subsample (1/4 of the total sample) to investigate the behaviour of
the spectral slope $\Gamma$.  The normalized distribution of spectral
slopes for the X--ray bright sample is shown in Figure \ref{g_histo}.
The distribution has been obtained by extracting the value of $\Gamma$
of each source $10^4$ times from the range allowed by the statistical
error bars, assuming a gaussian error distribution.  With this
procedure, we weight each source in the histogram according to the
statistical errors on $\Gamma$.  Before computing the weighted mean
value, we exclude the two brightest sources in the sample (about
$10^4$ net counts each) which otherwise would dominate the statistics.
We find that the weighted mean value for the spectral slope of the
bright sample is $\langle \Gamma \rangle = 1.75 \pm 0.02$ (error bar
refers to 1 $\sigma$ uncertainty on the mean value).  While the
typical error on a single measure is about $\Delta\Gamma \simeq 0.13$,
the dispersion of the distribution of the best fit values is $\sigma
\simeq 0.33$.  Assuming that both statistical errors and the intrinsic
dispersion in $\Gamma$ are distributed as a Gaussian, the intrinsic
scatter is of the order of $\sigma_{int} \sim 0.30$.  If we focus on
the 30 brightest sources to decrease the statistical errors (still
excluding the two sources with $\sim 10^4$ counts), the estimate of
the intrinsic scatter decrease to $\sigma_{int} \sim 0.20$, and the
weighted mean value is $\langle \Gamma \rangle = 1.81 \pm 0.01$.

\begin{figure}
\resizebox{\hsize}{!}{\includegraphics{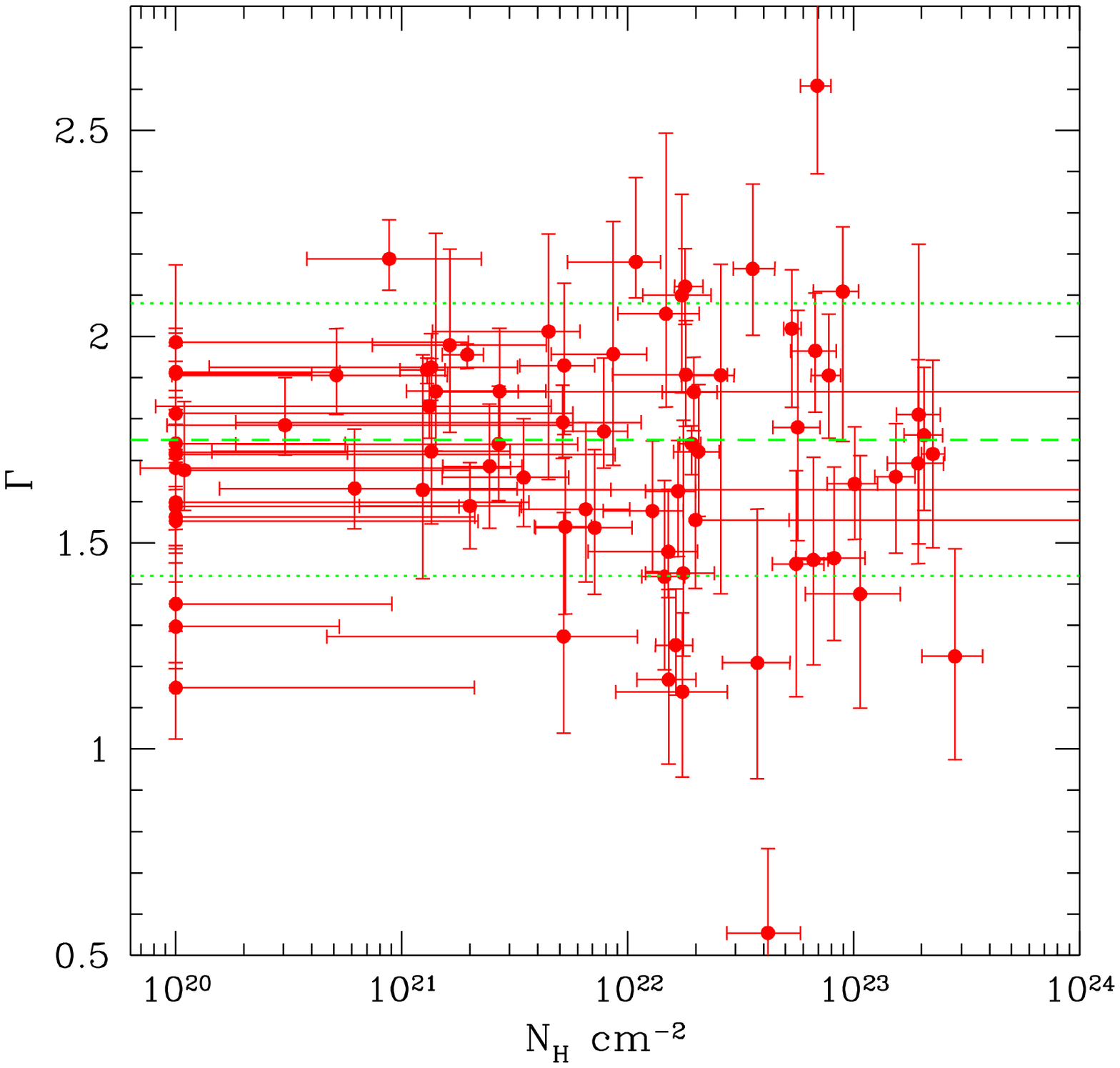}}
\caption{Scatter plot of the best fit values of $\Gamma$ and $N_H$ for
the bright X--ray sample (82 sources). Error bars correspond to 1
$\sigma$.  The dashed and dotted horizontal lines show the average
value of $\Gamma$ and its {\sl rms} dispersion respectively.
\label{nh_g_best}}
\end{figure}

\begin{figure}
\resizebox{\hsize}{!}{\includegraphics{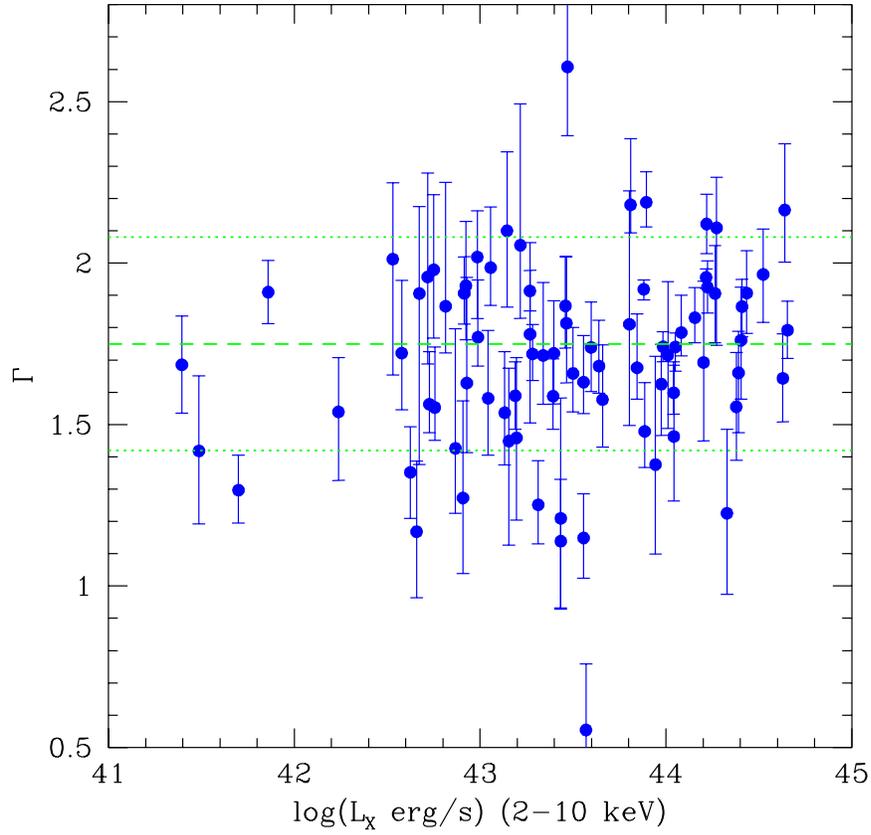}}
\caption{Scatter plot of the best fit values of $\Gamma$ versus
unabsorbed hard rest--frame luminosities for the bright X--ray sample
(82 sources).  Error bars correspond to 1 $\sigma$.  The dashed and
dotted horizontal lines show the average value of $\Gamma$ and its
{\sl rms} dispersion respectively.
\label{g_vs_l}}
\end{figure}

\begin{figure}
\resizebox{\hsize}{!}{\includegraphics{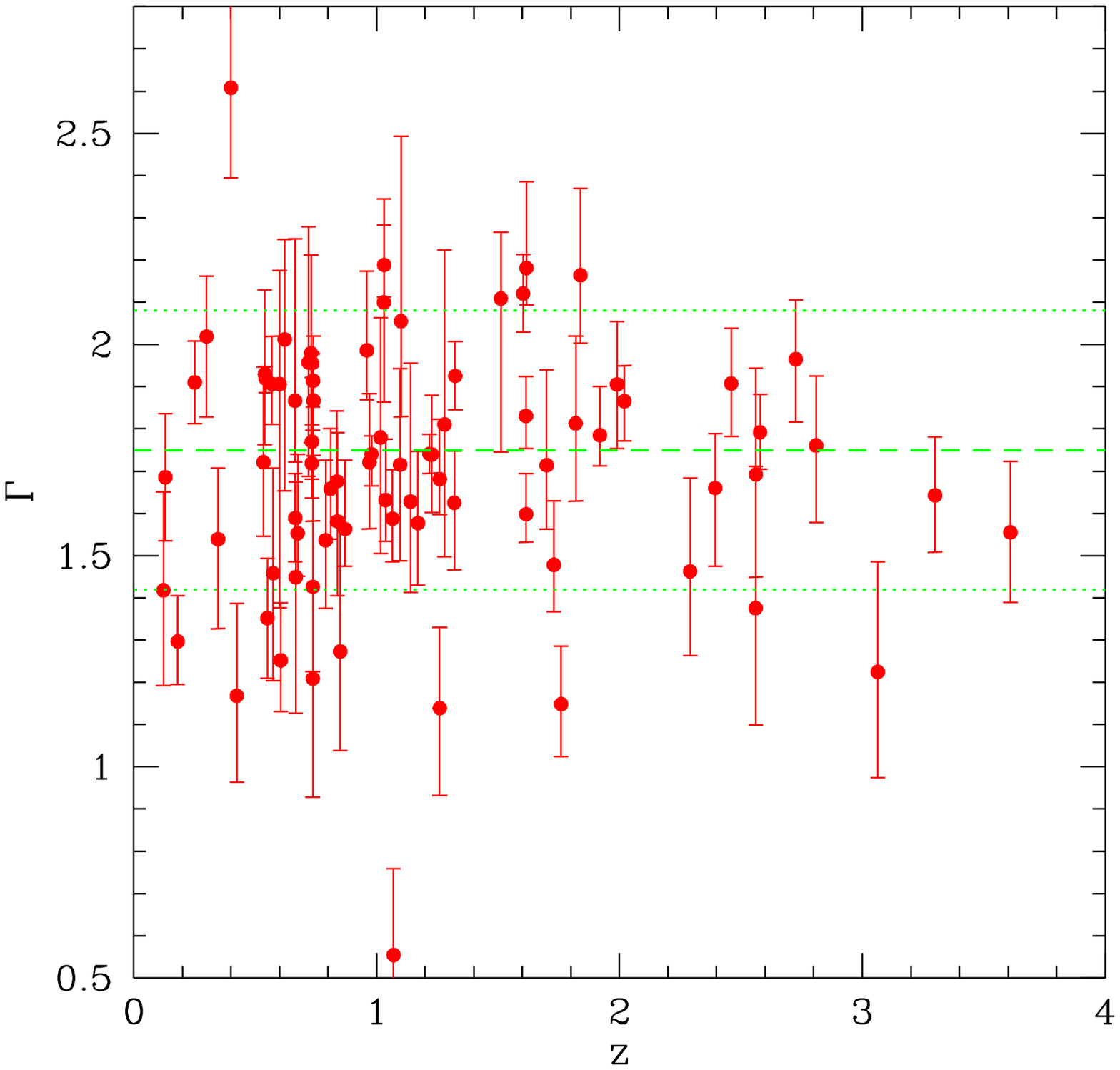}}
\caption{Scatter plot of the best fit values of $\Gamma$ versus
redshift for the bright X--ray sample (82 sources).  Error bars
correspond to 1 $\sigma$.  The dashed and dotted horizontal lines show
the average value of $\Gamma$ and its {\sl rms} dispersion
respectively.
\label{g_vs_z}}
\end{figure}

In Figure \ref{nh_g_best}, we plot the best fit values of $\Gamma$
versus the best fit values of the intrinsic absorption $N_H$.  We do
not detect any correlation between $\Gamma$ and $N_H$ (Spearman Rank
coefficient $SR \sim -0.04$). Note that if the intrinsic absorption is
close to the Galactic value for the CDFS field ($N_{Hgal} \simeq
8\times 10^{19}$ cm$^{-2}$) we are not able to derive any meaningful
value, due to the low--energy limit of our spectral range ($E >0.6$
keV). We considered these sources to be unabsorbed, plotting them at
$N_H=10^{20}$ cm$^{-2}$ in our Figures. We detect no correlation
between $\Gamma$ and the hard rest--frame intrinsic (unabsorbed)
luminosity (see Figure \ref{g_vs_l}).  The Spearman Rank correlation
is null also between $\Gamma$ and the redshift (see Figure
\ref{g_vs_z}).

From the analysis of the bright sample, we conclude that among our
sources the intrinsic continuum is well approximated by a power law
with $\Gamma\simeq 1.8$ (typical of Seyfert galaxies and AGN, as known
also from ASCA studies of AGN, see Turner et al. 1997) at any epoch.
On the other hand, it is well known that the flattening of the average
spectrum of the sources at low fluxes in deep X--ray survey is due
mainly to increasing intrinsic absorption (see Ueda et al. 1999b;
Tozzi et al. 2001; Piconcelli et al. 2003; La Franca et al. 2005).  In
addition, previous studies found no hints for a change in the slope of
the intrinsic power law as a function of epoch or luminosity (see also
Mainieri et al. 2002; Piconcelli et al. 2003; Vignali et al. 2003).
We conclude that the slope of the intrinsic power law can be assumed
to be constant for all the AGN population, and, therefore, we choose
to fix the spectral slope to $\Gamma=1.8$ when fitting the remaining
fainter sources, focusing on the $N_H$ distribution for the whole
sample.

\section{Results for the complete sample}

\begin{figure}
\resizebox{\hsize}{!}{\includegraphics{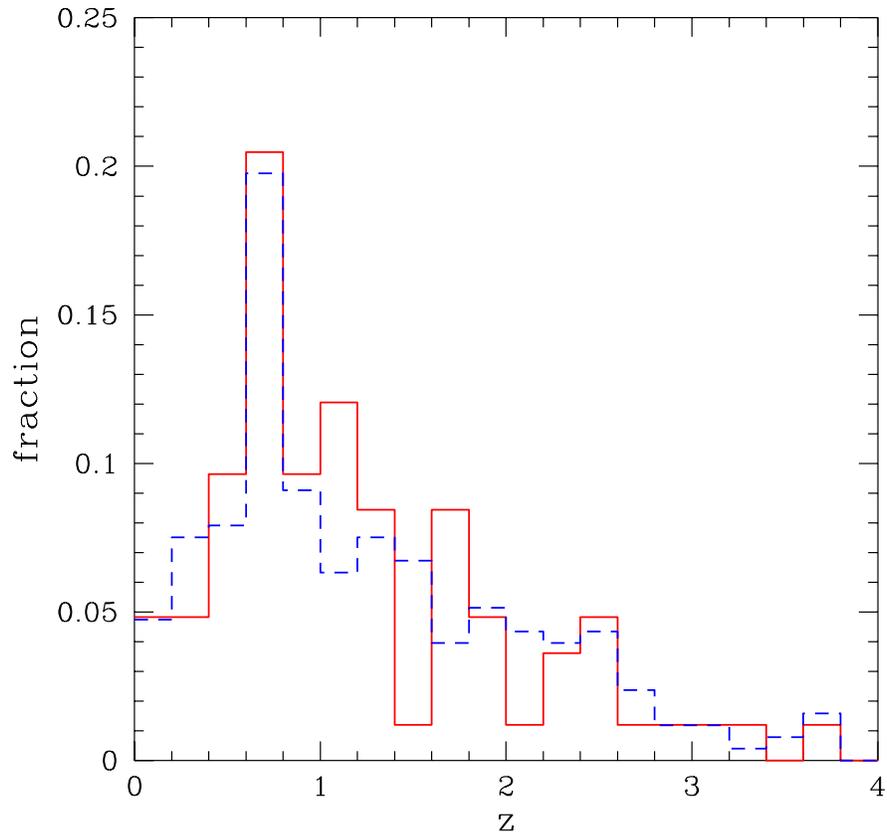}}
\caption{Normalized redshift distribution for the X--ray bright (solid
line, 82 sources) and the X--ray faint (dashed line, 253 sources)
subsamples.
\label{zhist}}
\end{figure}

We complete the analysis of the total sample fixing $\Gamma=1.8$ and
deriving $N_H$ for the remaining faint sources (239/321).  We remark
that our division in a bright and a faint subsample does not
correspond to a dramatic selection in redshift.  Indeed, the X--ray
bright and the X--ray faint subsamples have a similar distribution in
redshift (see Figure \ref{zhist}).  The results of the fits, along
with the redshifts and the quality of the optical spectra, are shown
in Table \ref{results}.

The distribution of the absorbing column densities is shown for the
whole sample in Figure \ref{nh_histo_mc}.  Our results are in good
agreement with preliminary results from the CDFN (Bauer et al. 2004a).
The distribution has been obtained by extracting the value of $N_H$ of
each source $10^4$ times from the range allowed by the statistical
error bars, assuming gaussian errors.  When the lower $\sigma$ error
bars hit zero, we adopt the upper error bar to allow the $N_H$
resampled value to go below zero; in this case, the resampled values
are included in the lowest bin.  The lowest bin shown is the value of
the Galactic absorption, $N_H \simeq 10^{20}$ cm$^{-2}$, below which
we cannot measure the intrinsic absorption, especially at high
redshifts.  This bin includes all the sources with nominal $N_H$ best
fit value lower than $10^{20}$ cm$^{-2}$.  Among these sources we
expect both redshifted AGN with low absorbing columns and normal
X--ray galaxies.  Note that here $N_H$ is an equivalent hydrogen
column measured assuming the photo--electric cross--sections by
Morrison \& McCammon (1983), with metal abundances relative to
Hydrogen by Anders \& Ebihara (1982).  The last bin at $N_H=10^{24}$
cm$^{-2}$ includes the few sources with measured $N_H> 10^{24}$
cm$^{-2}$ and the Compton--thick candidates.

We look for the Fe line only in those sources having at least 10 net
counts in both bands, to have an acceptable estimate of the continuum
and avoid spurious measures of high equivalent widths. Adopting a
threshold $\Delta C \geq 2.7$ with respect to the fit without the
line, corresponding to a minimum 90\% c.l. for one interesting
parameter, we find evidence for a significant Fe line in 20 sources
with at least 10 net counts in both bands.  The corresponding
equivalent widths span the 100-3000 eV range.  We carefully checked
that our criterion $\Delta C > 2.7$ actually corresponds to more than
90\% c.l. also in the case of a line detection (for which the
canonical confidence level criterion cannot be applied, see Protassov
et al. 2002). For each X-ray source we simulated 500 spectra starting
from the observed best fit model without the line. We then fitted each
simulated spectrum and looked for any variation in the C-stat when
adding a Fe line at $6.4/(1+z)$. The frequency of occurrence of
$\Delta C_{sim} > \Delta C_{obs}$ gives the probability $P$ that the
detected line is a statistical fluctuation.  In Figure
\ref{deltac_line} we show the significance (1-$P$) of the Fe line
versus the measured $\Delta C$.  We conclude that in the large
majority of the cases the criterion $\Delta C > 2.7$ corresponds to a
confidence level greater than 95\%.  Among the sources with more than
10 counts in both bands and a significant Fe line, 14/116 ($\sim
12$\%) are found among the sources with spectroscopic redshift, and
only 6/125 ($\sim 5$\%) are found in the subsample with photometric
redshift.  This shows that, given our X--ray spectral resolution, the
uncertainties in the photometric redshifts are likely to negatively
affect the detection of the Fe line with our method, i.e., fixing the
expected observing--frame energy of the line.  Indeed, we notice that
some sources do show strong hints of a Fe line at a redshift different
from the photometric one (see Mainieri et al. 2005a), or peculiar
lines (see Wang et al. 2003); finally, source variability could hide
the emission line (see Braito et al. 2005).  Therefore, we conclude
that the fraction of sources with significant emission line is
slightly larger than that found in an X--ray bright subsample in the
CDFN (7\%, see Bauer et al. 2004b).  In principle, if the Fe line were
produced only by the interaction of photons with the absorbing medium,
a positive correlation between $N_H$ and equivalent width might be
expected in obscured sources (Leahy \& Creighton 1993; Ghisellini,
Haardt \& Matt 1994). As shown in Fig.\ref{eqw}, we do not find strong
evidence of a correlation given the scatter of our data points, as
already observed (see Mushotzky, Done \& Pounds 1993).  The Fe lines
measured with low intrinsic absorption ($N_H <10^{22}$ cm$^{-2}$), may
be produced by the accretion disk, therefore breaking the expected
correlation.

In figure \ref{nh_vs_z} we show the scatter plot of intrinsic
absorption as a function of redshift for the whole sample.  We note
the lack of sources with high absorption ($N_H > 10^{22}$ cm$^{-2}$)
at $z<1$.  This is due to the fact that the low--luminosity, low--z
sources with high absorption show a strongly suppressed flux, and only
the intrinsically more luminous, rarer sources can be detected for a
given threshold in count rate; the detection probability, then,
decreases due to the small volume probed at low--z.  We also note a
lack of sources with low absorption (around $N_H \sim 10^{21}$
cm$^{-2}$) at high z.  This effect may be due to the difficulty in
measuring $N_H$ at $z>2$, since the absorption cutoff is redshifted
below the lower limit of the {\it Chandra} energy band we use (0.6
keV). This effect could result in spuriously high values of $N_H$ with
large error bars.  Note, however, that some of the points are just 1
$\sigma$ upper limits, implying the presence of sources with low $N_H$
value at high redshift as well.  It is clear that the $N_H$--$z$
scatter plot shows the effects of the incompleteness and partial
sampling of the AGN population.  Before investigating the shape and
evolution of the intrinsic $N_H$ distribution, we must correct for the
number of sources with a given $L_X$, $N_H$ and $z$ that fall outside
our detection criteria.  We will do this in the next Section.

In Figure \ref{nh_vs_l} we show the scatter plot of $N_H$ versus the
intrinsic, unabsorbed luminosities in the soft and in the hard band.
We remark that the intrinsic luminosities are computed in the
rest--frame soft and hard bands setting to zero the intrinsic
absorption in the XSPEC best fit model; for the Compton--thick
candidates we measure the intrinsic luminosities using a power law
model with $\Gamma=1.8$ and normalization fixed to that of the best
fit {\tt pexrav} model.  With this assumption the emitted (reflected)
luminosity of the C--thick sources is always about 6\% of the
intrinsic one in the hard band (while only 0.2\% in the soft).  We
also note that this model may give a lower limit to the intrinsic
luminosity, since its assumes a maximally efficient reflection; the
intrinsic luminosity can be higher for lower refelection efficiency
(Ghisellini, Haardt \& Matt 1994).  The envelope at low luminosity and
high $N_H$ is due to the fact that our survey is flux limited.  The
luminosity lower limit at a given redshift is not sharp, for two
reasons: first, our survey is count--rate limited, and different
spectral shapes may correspond to different fluxes and luminosities
for the same count rates; second, the unabsorbed luminosities are
related to the observed fluxes by a correction that depends on the
measured $N_H$.  For a preliminary investigation of a correlation
between $N_H$ and intrinsic luminosity, we select two regions in
Fig.\ref{nh_vs_l}: i) $L_X>10^{43}$ erg s$^{-1}$ and $N_H < 10^{24}$
cm$^{-2}$; ii) $L_X>10^{42}$ erg s$^{-1}$ and $N_H < 10^{23}$
cm$^{-2}$.  In this way we try to minimize the effects due to the
flux--limited nature of our sample.  In the first case, we do not find
significant correlation between $N_H$ and hard luminosity (Spearmann
Rank coefficient $SR = 0.06$ for 154 sources).  In the second case as
well, we do not detect significant correlation between $N_H$ and hard
luminosity (Spearmann Rank coefficient $SR = 0.08$ from 184
sources). This result is not in disagreement with results obtained
from larger samples.  Indeed, in flux--limited samples, the dependence
of the absorbed fraction on luminosity tends to be much weaker, as
discussed by Perola et al. (2004).  In the following, we will not
introduce by hand the correlation between the absorbed fraction and
luminosity found in larger sample spanning more than six decades in
flux.  The inability of retrieving in our sample such a correlation,
will not affect our main results, like the intrinsic distribution of
$N_H$, with the caveat that we are probing the luminosity range up to
few $\times 10^{44}$ erg s$^{-1}$.

In Figure \ref{nh_vs_l} we also show the locus of TypeII QSO, which is
the upper right corner marked with the dashed lines.  The criterion is
$L_X> 10^{44}$ erg s$^{-1}$ and $N_H > 10^{22}$ cm$^{-2}$.  For a
spectral slope of $\Gamma = 1.8$, a total luminosity of $10^{44}$ erg
s$^{-1}$ in the 0.5-10 keV band corresponds to $3.9 \times 10^{43}$
erg s$^{-1}$ in the 0.5--2 keV band and $6.1 \times 10^{43}$ erg
s$^{-1}$ in the 2--10 keV band.  With these criteria, using X--ray
spectral parameters and, most importantly, unabsorbed luminosities,
the number of QSOII in the CDFS sample is 54.  This corresponds to a
surface density of X--ray selected QSO equal to $(620 \pm 80)$ sq
deg$^{-2}$ at the flux limit of $5 \times 10^{-16}$ erg cm$^{-2}$
s$^{-1}$.  This is higher than the value found by Padovani et
al. (2004), but the difference is due to their selection based on the
condition $L_{2-10}> 10^{44}$ erg s$^{-1}$.  Applying the same
criteria, we find a surface density of $(360 \pm 50)$ sq deg$^{-2}$ in
very good agreement with Padovani et al. (2004; see also La Franca et
al. 2005).  We note, however, that the density of TypeII QSO depends
sensitively on the luminosity cut in the intrinsic power used in the
analysis.

Finally, we present a sample of 14 Compton--thick candidates selected
only on the basis of the X--ray spectral shape with the selection
thresholds described in \S3.2.  Two of them were already identified as
Compton--thick sources on the basis of multiwavelength data (source ID
202 and 263, see Norman et al. 2002; Mainieri et al. 2005b).  We
assign a value $N_H \gtrsim 1.5 \times 10^{24}$ cm$^{-2}$ to our
Compton--thick candidates.  Among them, 2 sources (out of 7 with
secure spectroscopic redshift) show a Fe K emission line, while no
Compton thick candidate source with photometric redshift does show a
statistically significant line.  We believe that the uncertainties in
the photometric redshift prevent us from recovering the line.  We also
note that some high column density sources at low redshift may not
have strong Fe K lines (see Fruscione et al. 2005).  We checked that
the distribution of the net detected counts of the C--thick candidates
is not different from that of the whole sample, indicating that there
are no evident bias due to the low signal--to--noise. The
net--detected counts for the C--thick sample ranges from 170 to 40,
with an average of 65.  We notice that for these sources the detection
probability is low, due to their hard spectra.  Consequently, their
associated sky--coverage is low, and their surface density
correspondingly higher, close to $(200 \pm 50)$ deg$^{-2}$.  The
actual surface density of C--thick sources may be 20\% higher if
including selection effects (see Appendix C).  We notice also that the
fraction of C--thick sources predicted by updated models for the
synthesis of the XRB is in very good agreement with that found in the
CDFS (Gilli, Comastri \& Hasinger 2006, in preparation).

\section{Intrinsic absorption distribution and its 
evolution with cosmic epoch}

\begin{figure}
\resizebox{\hsize}{!}{\includegraphics{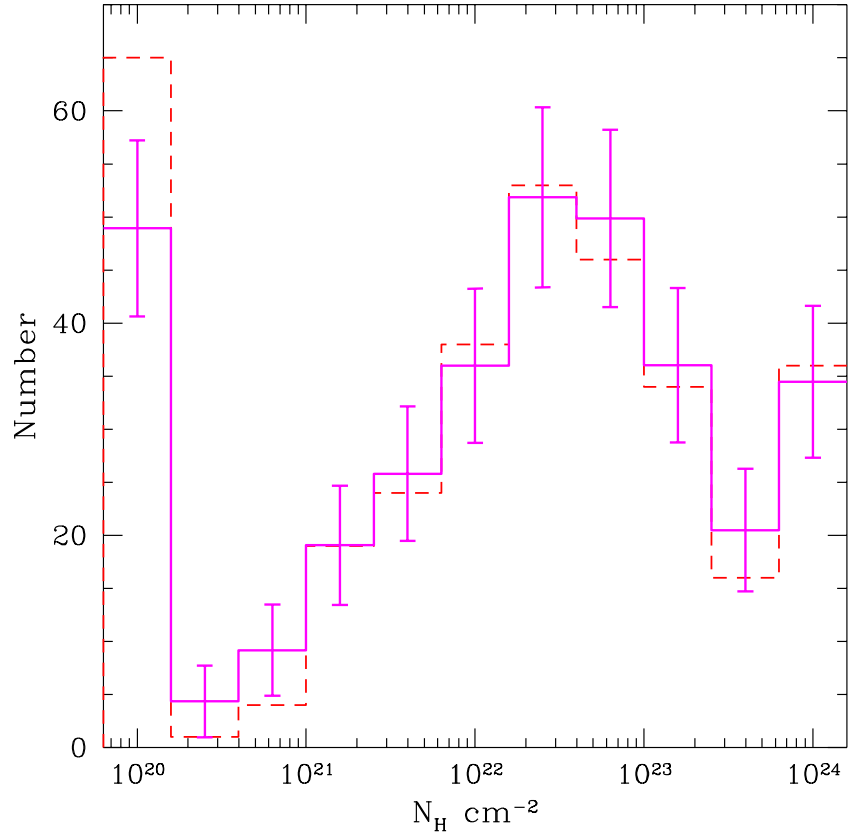}}
\caption{The solid line shows the observed $N_H$ distribution for the
whole sample (321 sources).  Error bars refer to 1$\sigma$ poissonian
uncertainty due to the limited number of sources in each bin.  The
dashed histogram shows the distribution obtained without resampling
according to measurement statistical errors.  No correction for
incompleteness and volume--sampling effects has been applied.
\label{nh_histo_mc}}
\end{figure}

\begin{figure}
\resizebox{\hsize}{!}{\includegraphics{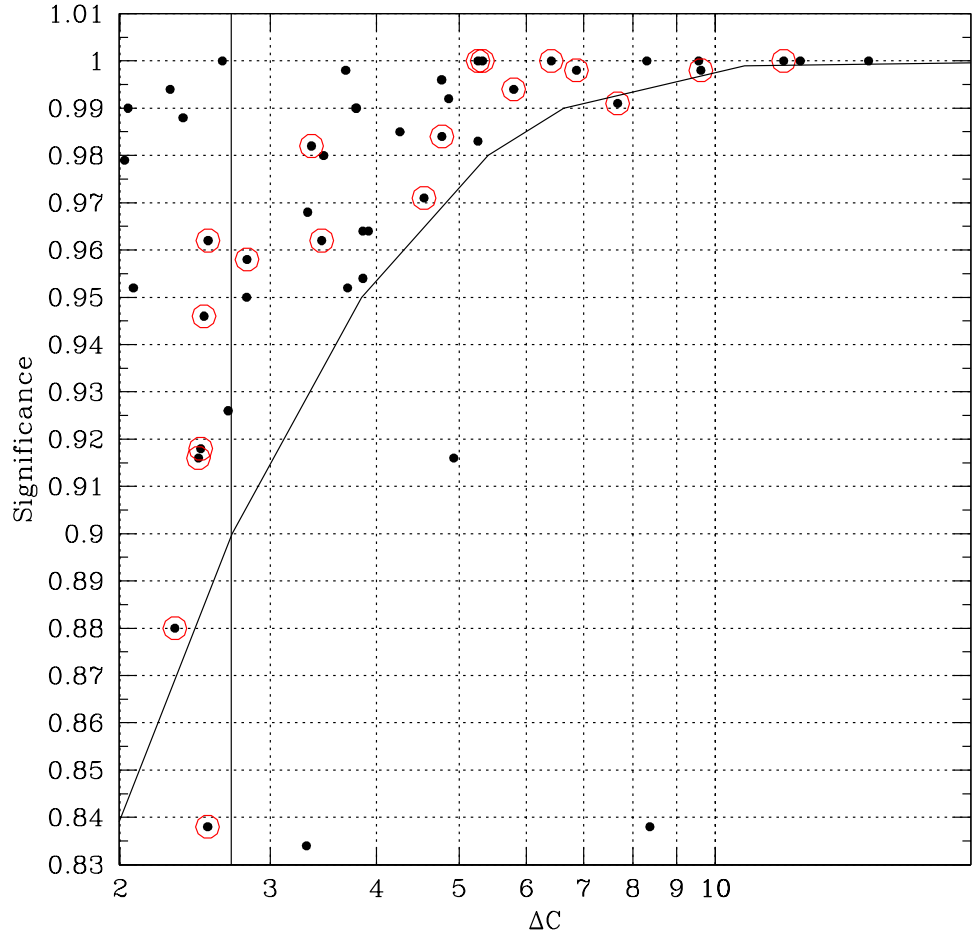}}
\caption{Significance of the Fe line (measured with simulations, see
text) plotted versus the observed $\Delta C$.  Pointed circles are
sources included in the bright sample.  The solid line shows the
significance vs $\Delta C$ for one interesting parameter assuming the
same statistics for $\Delta C$ and $\chi^2$.  The vertical line is the
threshold corresponding to $\Delta C=2.7$.
\label{deltac_line}}
\end{figure}

\begin{figure}
\resizebox{\hsize}{!}{\includegraphics{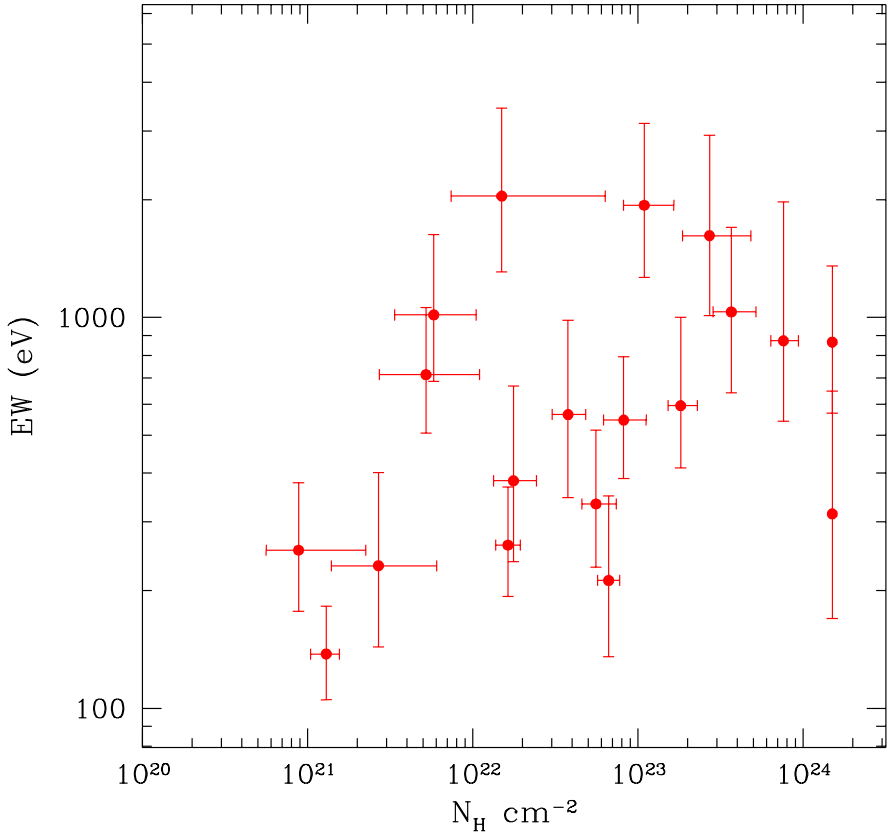}}
\caption{Equivalent width of the Fe line plotted versus the intrinsic
absorption $N_H$ for the 20 sources with Fe line significant at more
than 90\% c.l.  Errors on the equivalent width are derived from the
errors on the normalization of the line component.  Compton--thick
candidates are plotted at $N_H = 1.5\times 10^{24}$ as lower limits to
the actual value.
\label{eqw}}
\end{figure}

\begin{figure}
\resizebox{\hsize}{!}{\includegraphics{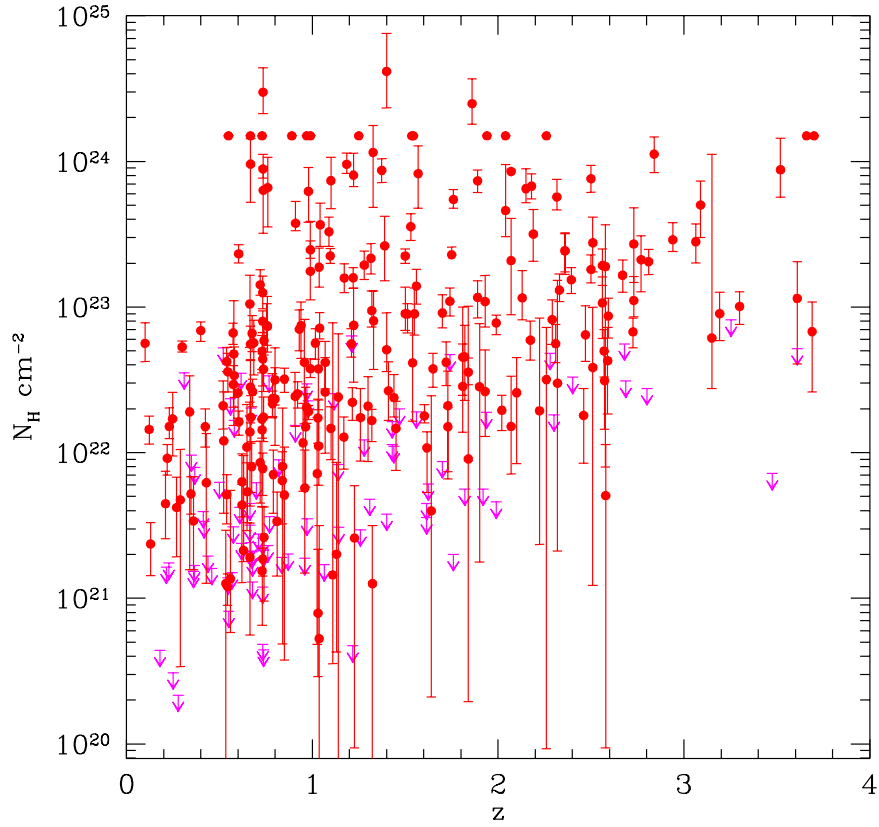}}
\caption{Intrinsic absorption versus redshift for the complete sample.
Upper limits (1 $\sigma$) are used for measures consistent with
$N_H = 0$ within 1 $\sigma$.  Compton--thick candidates are plotted at
$N_H = 1.5\times 10^{24}$ as lower limits to the actual value.  Error
bars correspond to 1 $\sigma$.
\label{nh_vs_z}}
\end{figure}

\begin{figure}
\resizebox{\hsize}{!}{\includegraphics{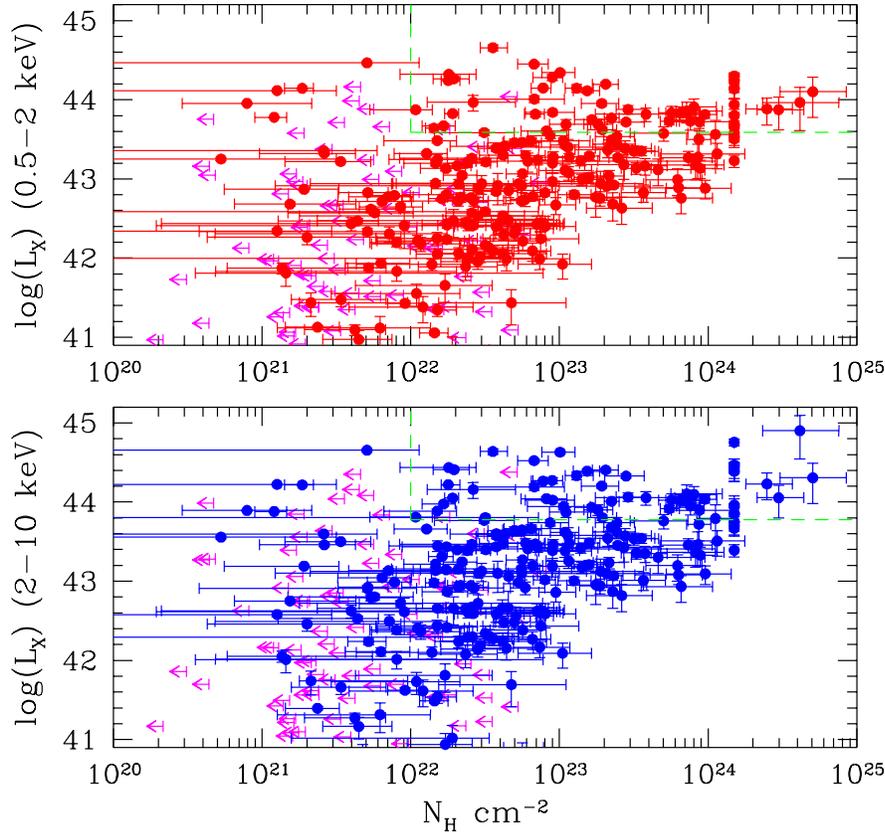}}
\caption{Unabsorbed rest--frame luminosities in the soft (upper panel)
and in the hard (lower panel) band plotted versus the intrinsic
absorption.  Upper limits (1 $\sigma$) are used for for measures
consistent with $N_H = 0$ within 1 $\sigma$.  Error bars correspond to
1 $\sigma$.  Upper right corners outlined by the dashed lines show the
locus of QSO--II, defined as sources with $L_X > 10^{44}$ erg s$^{-1}$
and $N_H> 10^{22}$ cm$^{-2}$ (as opposed to the criterion $HR>-0.2$
and $L_X>10^{44}$ erg s$^{-1}$ used in Szokoly et al. 2004).
\label{nh_vs_l}}
\end{figure}

In this Section, we estimate the intrinsic absorption distribution
(the $N_H$ function) for the AGN population in our sample.  The
distribution of $N_H$ that we showed in Figure \ref{nh_histo_mc}, does
not include any correction for incompleteness, and it refers only to
the sources observed in the region of the $N_H$--$L_X$--$z$ space
which is delimited by the count--rate detection thresholds of the
survey.  To go from this distribution to a distribution which is
representative of the whole AGN population, we must apply two
independent corrections.  The first is the completeness correction and
it is given by the effective solid angle under which a source of a
given intrinsic luminosity, absorbing column density and redshift, is
detected in the CDFS with our criteria.  The second correction takes
into account the sources which are outside the detectability region in
the $N_H$--$L_X$--$z$ space, and therefore it must be based on a
specific model of the luminosity function of AGN.  We remind that a
reliable luminosity function cannot be obtained from CDFS data alone,
but should rather be derived from a combination of wider surveys, in
order to sample the bright end of the luminosity distribution, which
is poorly represented in our pencil beam survey (see Brandt \&
Hasinger 2005).  We describe these two corrections below.

To correct for incompleteness, we simply weight each source for the
inverse of the solid angle under which the source can be detected in
the CDFS.  To measure this quantity, first we compute the net count
rate in the soft and hard band that would be measured in the aimpoint
of the CDFS for each source in the sample, using its best--fit model.
Then, we measure the solid angle $\omega_i$ where the $i$th source can
be detected in the CDFS, including the vignetting correction and the
background evaluated locally.  Since the detection threshold is
applied separately in the hard and the soft image, the effective solid
angle is the largest between the two. We recall that our survey is
limited in count rate, not in flux, and for a given intrinsic
luminosity and redshift, the count rate is strongly dependent on the
intrinsic absorption, especially in the soft band, where the
sensitivity of our survey is the highest.  Most of the sources have
the largest detectability angle in the soft band, while the fewer,
strongly absorbed, hard sources have the largest detectability solid
angle in the hard image.  The {\sl a priori} probability of having a
given source included in the CDFS sample is simply the ratio of the
solid angle $\omega_i$ to the total solid angle covered by the 11
exposures of the CDFS ($\omega_{CDFS} = 0.108$ deg$^2$).  Then, when
binning our sample as a function of the measured $N_H$, we weight each
source for the inverse of its detection probability:

\begin{equation}
F(N_H) dN_H= \sum_{N_H bin} P_i \times (\omega_i /
\omega_{CDFS})^{-1}  \, .
\label{compl_corr}
\end{equation}

Here, the weight $P_i$ would be equal to 1 if $N_H$ were measured with
negligible error with respect to the size $dN_H$ of the bin.  To
account for statistical uncertainties in the measured value of $N_H$
for each source, we put $P_i$ equal to the probability that the actual
value falls within the $N_H$ bin, according to the best fit value and
its error bars.  The error on $F(N_H)$ is the poissonian error
associated to the number of sources counted in the bin $N_H$--$N_H+
dN_H$.

Then, we compute the second correction, to account for the sources
which are outside the detectability region in the $N_H$--$L_X$--$z$
space in the CDFS survey.  This correction is relevant for strongly
absorbed sources, since our limit in count--rate allows us to sample a
smaller range of intrinsic luminosity for increasing $N_H$ at a given
redshift.  This effect is mitigated at high redshift due to the
positive X--ray K--correction.  Therefore, for any given redshift and
luminosity, we are measuring a different fraction of unabsorbed and
absorbed sources with respect to the total AGN population.  As a
consequence, the directly observed fraction of sources with a given
$N_H$ is affected by the shape of the actual AGN luminosity function
and by its cosmic evolution.

To correct for this effect, we must assume a model for the AGN
luminosity function.  One of the most recent is the Luminosity
Dependent Density Evolution model obtained by Ueda et al. (2003; but
see Barger et al. 2005 for another determination of the AGN X--ray
luminosity function consistent with pure luminosity evolution), in
which low--luminosity sources peak at lower redshift than
high--luminosity AGN. Such a luminosity function is measured from a
combination of surveys with HEAO--1, ASCA and Chandra including part
of the CDFN sample (see also Hasinger, Miyaji and Schmidt 2005 for the
most recent measure of the Type I AGN luminosity function).  In
particular, we use equations 11-15-16-17 of Ueda et al. (2003) to
write the comoving density of AGN per hard--band luminosity interval
$N(L_X,z)$.

After assuming a luminosity function for the whole AGN population, we
can write the number of AGN in a given interval of $N_H$, $L_X$ and
$z$ as

\begin{equation}
F(N_H, L_X, z)dN_H dL_X dz = N(L_X,z) dL_X {{dV}\over{dz}}
dz \,\,\,\, f(N_H, L_X, z) dN_H \, , 
\end{equation}

\noindent
where $V$ is the comoving volume element, and $f(N_H,L,z)$ is the
probability of measuring an intrinsic absorption between $N_H$ and
$N_H+dN_H$ for a given $L_X$ and $z$.  Let's assume that $f(N_H,L,z)$
is slowly varying as a function of $L_X$ and $z$ in our sample.  The
total number of sources that we are detecting in our survey with
intrinsic absorption between $N_H$ and $N_H+dN_H$ is then given by:

\begin{equation}
F(N_H)dN_H = f(N_H) dN_H \int^{z_{max}}_0 {{dV}\over{dz}} dz
\int^{L_{max}}_{L_{cut}(N_H,z)} N(L_X,z) dL_X \, \,
\end{equation}

\begin{figure}
\resizebox{\hsize}{!}{\includegraphics{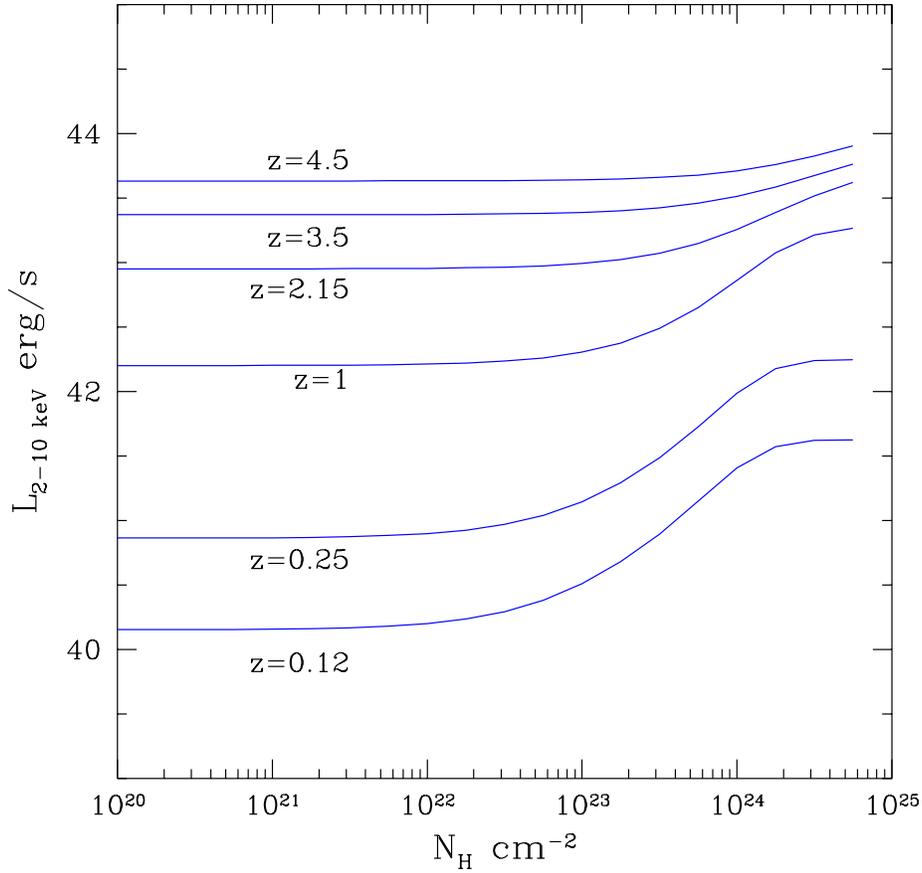}}
\caption{Intrinsic minimum rest--frame luminosity in the 2--10 keV
band ($L_{cut}$) of a source that can be detected at the CDFS aimpoint
as a function of intrinsic absorption for redshifts
$z=0.12-0.25-1-2.15-3.5-4.5$ from the bottom to the top.  The assumed
model is a Compton thin power law with $\Gamma = 1.8$ plus a
reflection component equal to 6\% of the hard intrinsic luminosity.
\label{lcut}}
\end{figure}

\noindent
Here the luminosity $L_{cut}(N_H,z)$ is the 2--10 keV intrinsic
luminosity for which, at any given z and $N_H$, the net count rate is
equal to the minimal count rates in the hard {\it or} in the soft
band.  The minimal count rates for detection in the aimpoint of the
CDFS are $1.2\times 10^{-5}$ cts/s in the soft and $1.5\times 10^{-5}$
cts/s in the hard band.  These values are defined with small
uncertainties because of the rapid drop of the sky coverage as a
function of the count rate in both bands.  To compute $L_{cut}$, we
assume that in average our sources can be described with a
Compton--thin model with spectral slope fixed to $\Gamma =1.8$, plus a
reflection component with the same slope and normalization.  The
reflection component (modeled with the {\tt pexrav} XSPEC model)
amounts to 6\% of the hard intrinsic luminosity.  Such a reflection
component will dominate the emission of the Compton--thick sources
with $N_H \geq 1.5 \times 10^{24}$ cm$^{-2}$.  The value of $L_{cut}$
as a function of $N_H$ is shown in Figure \ref{lcut} for different
redshifts.  We note that for unabsorbed sources ($N_H < 10^{22}$
cm$^{-2}$) the cut depends only on the intrinsic luminosity at any
redshift.  However, for larger column densities, the cut in luminosity
is higher for larger $N_H$, but the effect is weaker at higher $z$
where the positive X--ray K--correction shifts the hard rest--frame
emission in the soft band.  In the Compton--thick regime, a roughly
constant fraction of the intrinsic luminosity reflection by cold
material dominates the emission, making $L_{cut}$ flat again.  We do
not attempt to include the effect of the presence of the scattered
component, which is detected only in less than 3\% of the sources in
our sample.

\begin{figure}
\resizebox{\hsize}{!}{\includegraphics{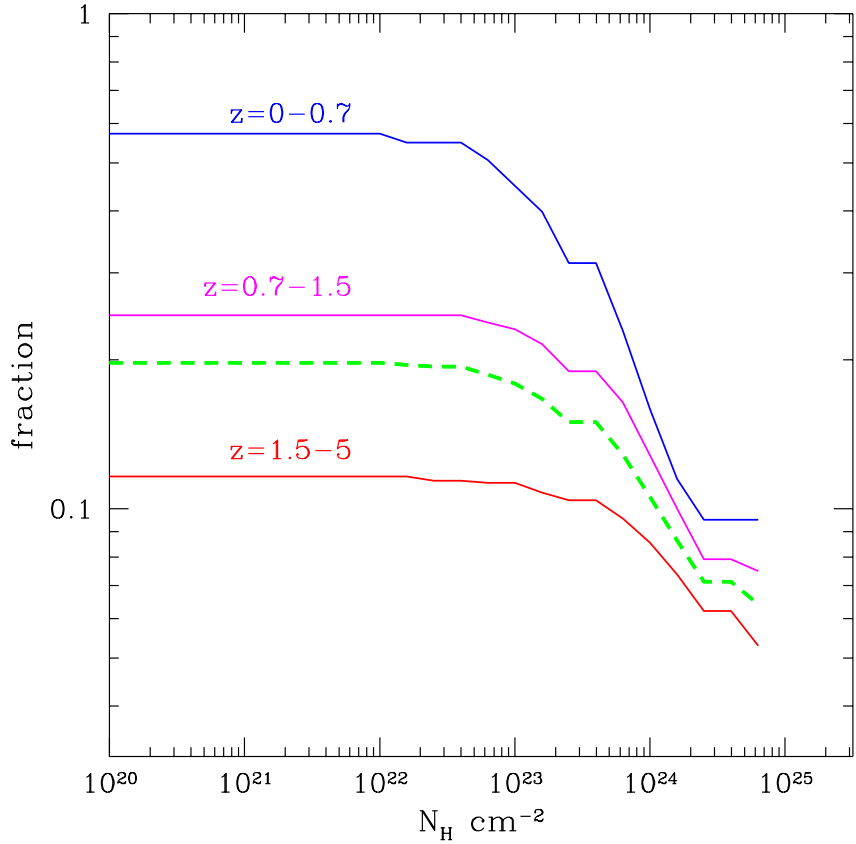}}
\caption{ Total fraction of all the AGN detected in the CDFS survey as
a function of the intrinsic $N_H$, according to the detection criteria
in the CDFS and assuming the luminosity function of Ueda et
al. (2003).  The total fraction strongly depends on the minimum
detectable luminosity and hence, given the flux limit in the CDFS, on
the redshift range.  Note that low fractions are implied by the
conservatively low value $L_{min}$ ($10^{41}$ ergs s$^{-1}$) which
defines the total population of AGN.  The thick, dashed line is the
correction for the whole sample.  The three continuous lines refer to
three intervals in redshifts: $z=0-0.7$, $z=0.7-1.5$, $z=1.5-5$ from
top to bottom.
\label{nh_corr}}
\end{figure}

Since $F(N_H)$ (computed with equation 1) is the directly observed
$N_H$ distribution (after correcting for incompleteness), the
probability function $f(N_H)$ can be obtained after equation 3
(discretizing the integral over $N_H$).  The resulting fraction of AGN
visible in the CDFS as a function of $N_H$ is shown in Figure
\ref{nh_corr} for three different redshift intervals (solid lines),
and for the whole explored redshift range (thick dashed line).  This
fraction is computed as the ratio of the detectable AGN over the total
number of AGN predicted by the Ueda et al. luminosity function in the
range $L_{max} = 10^{45}$ ergs s$^{-1}$, $L_{min} = 10^{41}$ ergs
s$^{-1}$, and $z_{max} = 5$.  Note that the low values of this
fraction does not imply that the majority of the AGN are not detected
in the CDFS; in fact, such low values are mostly due to the
conservatively low minimum luminosity adopted here ($L_{min} =
10^{41}$ ergs s$^{-1}$) and depend on the faint end slope of the
luminosity function.  These aspects, in turn, weakly affects the
dependence of the fraction on $N_H$, which is our main concern here.
Here we do not discuss the effects of the shape of the underlying
luminosity function, postponing this to a subsequent paper.
Therefore, we estimate in a robust way the dependence of the total
fraction of visible AGN on the redshift (given the flux limit in the
CDFS) and on $N_H$.  The fraction decreases towards higher values of
$N_H$ due to the reduced emission in the soft band, but it flattens
again in the Compton--thick regime, where the emitted luminosity is
roughly a constant fraction of the intrinsic one.

\begin{figure}
\resizebox{\hsize}{!}{\includegraphics{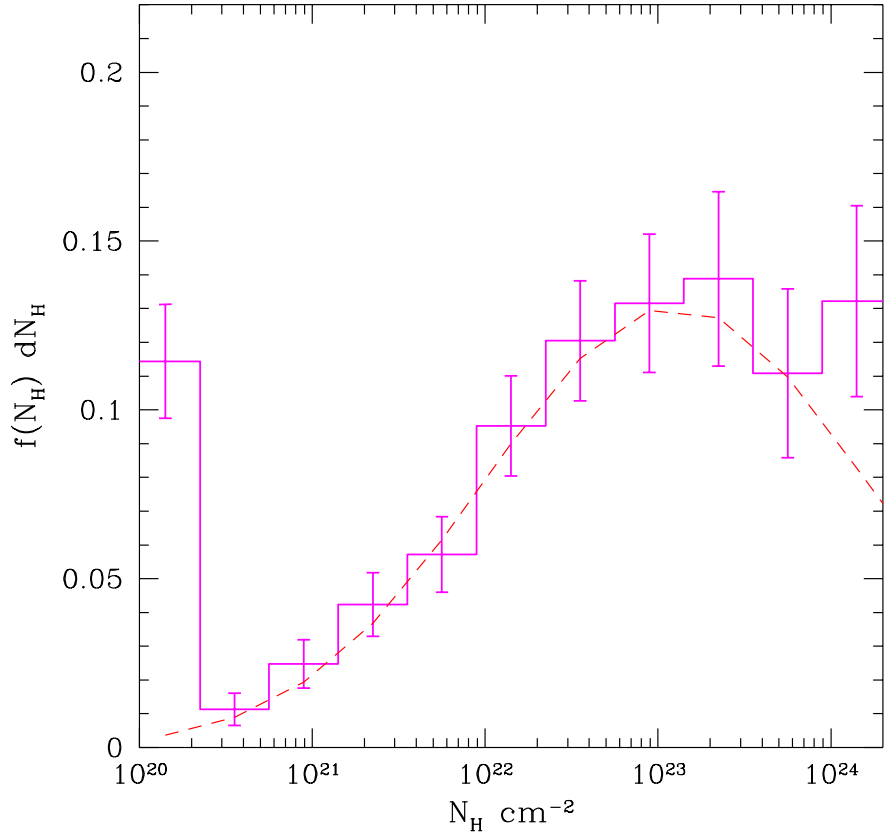}}
\caption{Intrinsic $N_H$ distribution representative of the whole AGN
population in our sample (corrected for incompleteness and
sampling--volume effect, and convolved with the statistical errors of
each measurement).  Errors are obtained from the poissonian
uncertainties on the number of detected sources in each bin.  The
dashed curve is a lognormal distribution with $\langle log(N_H)
\rangle = 23.1$ and $\sigma = 1.1$.  Compton--thick candidates are all
in the bin at $N_H = 10^{24}$.
\label{nh_histo_conv}}
\end{figure}

The corrected, normalized distribution of the intrinsic absorption for
the whole sample is shown in Figure \ref{nh_histo_conv}.  Errors are
obtained from the poissonian uncertainties on the number of detected
sources in each bin.  The distribution that we measured is bimodal, in
the sense that 10\% of sources have $N_H < 10^{20}$ cm$^{-2}$ and
appear separated from the distribution of the bulk of the sources.
However, we remark that the fraction of sources with negligible
absorption in our sample may include normal galaxies with high star
formation rate.  The distribution of the bulk of the sources can be
roughly approximated with a lognormal distribution centered on
$\langle log(N_H) \rangle \simeq 23.1$ and with a dispersion $\sigma =
1.1$.  We remark that in the Compton--thin regime, where our estimates
are more robust, the number of obscured sources is steeply increasing
with $N_H$ in agreement with Risaliti et al. (1999) and Dwelly et
al. (2005).

This distribution accounts for the Compton--thin sources with
intrinsic absorption up to $N_H \simeq 10^{24}$ cm$^{-2}$, and for
Compton--thick sources at higher absorption, bridging the bulk of the
AGN to the Compton--thick population.  This is the main difference
with the distribution presented in Treister et al. (2004), where the
fraction of sources with $N_H > 10^{23} $ cm$^{-2}$ is dropping.
Indeed, strongly absorbed AGN are expected to be missed by surveys
that rely on optical spectroscopy.  Here we show that part of the
population of Compton--thick sources can be detected in present deep
X--ray Surveys via a careful spectral analysis of all the X--ray
detected sources, avoiding selection based on optical spectroscopy.
Our results are consistent with the preliminary results on the $N_H$
distribution found in the CDFN (Bauer et al. 2004a), which already
shows a peak at larger $N_H$ values with respect to the results of
Ueda et al. (2003).  We remark that this result is not affected by
small variations with respect to the luminosity function proposed by
Ueda et al. (2003), which indeed is consistent with the present data
on the AGN luminosity distribution.  To summarize, we conclude that at
least part of the expected population of strongly absorbed AGN
(expected to be observed in the submillimiter with the {\it Spitzer}
satellite) is already present in the deep X--ray Survey such as the
CDFS.

\begin{figure}
\resizebox{\hsize}{!}{\includegraphics{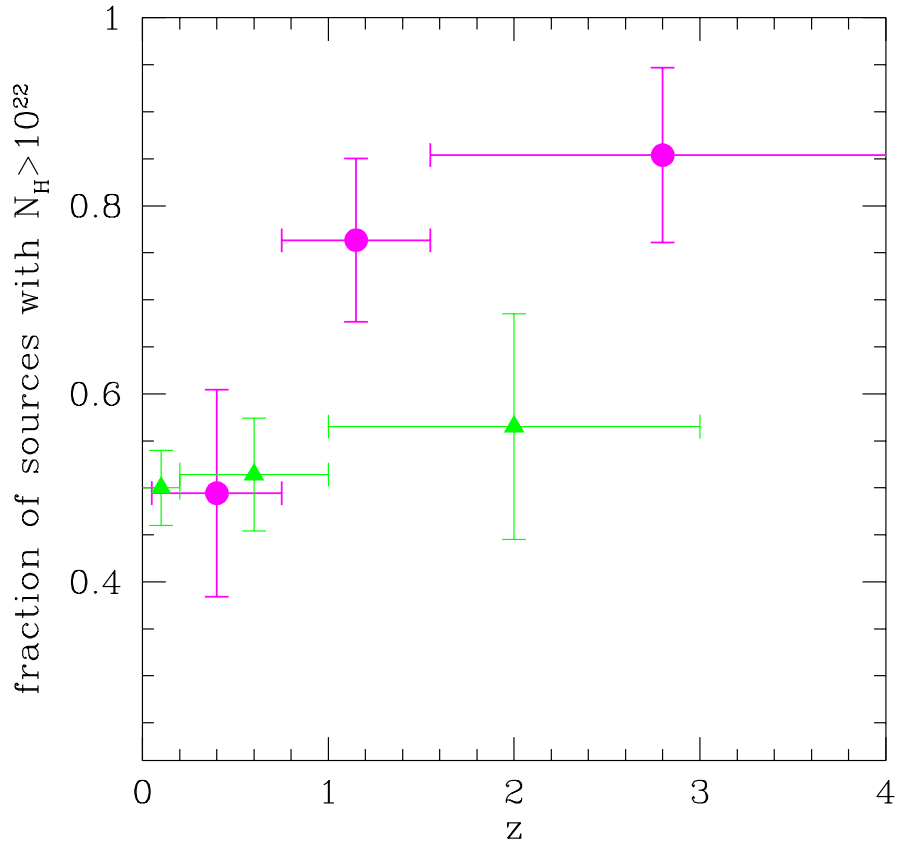}}
\caption{Fraction of absorbed AGNs with $N_H>10^{22}$ cm$^{-2}$ to all
AGNs with $L_X>10^{41}$ erg s$^{-1}$ (2--10 keV band) as a function of
redshift (solid circles).  Triangles are the data points from Ueda et
al. (2003) for $10^{43} < L_X < 10^{44.5}$ erg s$^{-1}$ (2--10 keV
band).  Rest--frame luminosities are computed for a $\Lambda = 0.7$
flat cosmology and $H_0 = 70$ km/s/Mpc.
\label{ratio_vs_z}}
\end{figure}

The $N_H$ function is derived under the assumption of no strong
intrinsic correlation between $L$ and $N_H$ or $z$ and $N_H$ in our
sample, so that we can obtain $f(N_H)$ without binning our sample as a
function of $L_X$ or $z$.  However, here we investigate for possible
evolution with redshift of the absorbed fraction of sources.  Due to
the limited statistics, we focus on the cosmic evolution of the ratio
of absorbed sources ($N_H>10^{22}$ cm$^{-2}$) to all the AGNs in three
bins of redshift.  The redshift bins are $z=0-0.7$, $0.7-1.5$,
$1.5-5$, including 76, 125 and 109 sources with $L_X>10^{41}$ erg
s$^{-1}$ respectively (the first two bins include the two most
prominent spikes in the CDFS redshift distribution at $z=0.67$ and
$z=0.73$, as shown in Gilli et a. 2003).  The correction for the
absorbed sources that are missed is larger at low redshift, as can be
seen in Figure \ref{nh_corr} (upper curve for the redshift range
$z=0$--$0.7$), while at high z is almost flat up to log($N_H$)$=23.5$
(lower curve for the redshift range $z=1.5$--$5$).  In Figure
\ref{ratio_vs_z} we show that the absorbed fraction is consistent with
a moderate increase, in agreement with the model of Gilli et
al. (2001; see also Civano, Comastri \& Brusa 2005).  We remark that
the absorbed fraction in the first bin at $z<0.8$, including the low
luminosity sources, may be underestimated due to the presence of star
forming galaxies in the luminosity range $10^{41}$--$10^{42}$ erg
s$^{-1}$.

We note that the overall value of the fraction of absorbed sources is
larger than that found by Ueda et al. (2003).  However, the points of
Ueda et al. (2003) include only sources with $L_X > 10^{43}$ erg
s$^{-1}$, and therefore are expected to be significantly higher when
including lower luminosities.  The global fraction of absorbed sources
is in agreement with that estimated in the CDFN by Perola et
al. (2004), and with a ratio of absorbed over unabsorbed sources in
the sample of about 4, as observed in the local Universe (eg. Maiolino
\& Rieke 1995).  This value is also consistent with the theoretical
expectation of 3/4 of all the AGN being absorbed as in the standard
unification scenario (Antonucci 1993).  While in the CDFS and CDFN
the fraction of obscured sources seems to be in agreement with the
expectations of the standard unification scenario and popular
synthesis models of the X--ray background, in shallower serendipitous
surveys like those performed with XMM by Piconcelli et al. (2003) and
Mateos et al. (2005) obscured sources seem to be a factor of $\simeq
2$ less abundant.  At typical X--ray fluxes of a few $10^{-14}$ erg
s$^{-1}$ cm$^{-2}$, XMM serendipitous sources have a median luminosity
of a few $10^{44}$ erg s$^{-1}$. It is therefore possible that the
intrinsic fraction of obscured sources is decreasing at luminosities
higher than that observed in the Chandra Msec fields, which would
point towards a paucity of obscured QSOs as found by Ueda et
al. (2003; see also La Franca et al. 2005).  Alternatively, one could
argue about the large spectroscopic incompleteness of XMM samples
(more than 60\% of the sources are as yet unidentified) before drawing
solid conclusions.

\section{Comparison between X--ray and Optical properties}

If we classify the whole sample of 321 sources with $L_X>10^{41}$ erg
s$^{-1}$, according only to the optical spectra, we obtain the
following:

\begin{itemize}
\item 34 Broad Line AGN (BLAGN);
\item 20 High Excitation Line galaxies (HEX);
\item 67 Low Excitation Line Galaxies (LEX);
\item 22 Absorption spectrum typical of late--type galaxies;
\item 178 non classified.
\end{itemize}

In this section we compare the optical classification with the X--ray
classification, to investigate if a revision of the unification model
is actually needed (see, e.g., Matt 2002).  This was already done in
Szokoly et al (2004); the main difference here is that we use
unabsorbed luminosities and intrinsic absorption as opposed to
absorbed luminosities and hardness ratio, providing therefore a more
physical X--ray classification.  We use the value $N_H = 10^{22}$
cm$^{-2}$ as the threshold to divide X--ray unabsorbed sources from
X--ray absorbed ones.  We define normal X--ray galaxies the sources
with $N_H<10^{22}$ cm$^{-2}$ and $L_X < 10^{42}$ erg s$^{-1}$.  Our
results are shown in Table \ref{optype}, to be compared with Table 8
of Szokoly et al. (2004).  We remark that the class ``normal
galaxies'', amounting to 42 sources, may include low luminosity AGN.
Indeed, if we restrict our criterion to source with low intrinsic
absorption (values $N_H<10^{21}$ cm$^{-2}$ can be due also to diffuse
matter in the host galaxy, as opposed to the larger absorbing columns
typical of circumnuclear matter), the normal galaxies class would
include 23 sources only.  Therefore, we can bracket the contamination
of our sample by normal galaxies to be between 7\% and 14\% of the
total sample. 

\begin{figure}
\resizebox{\hsize}{!}{\includegraphics{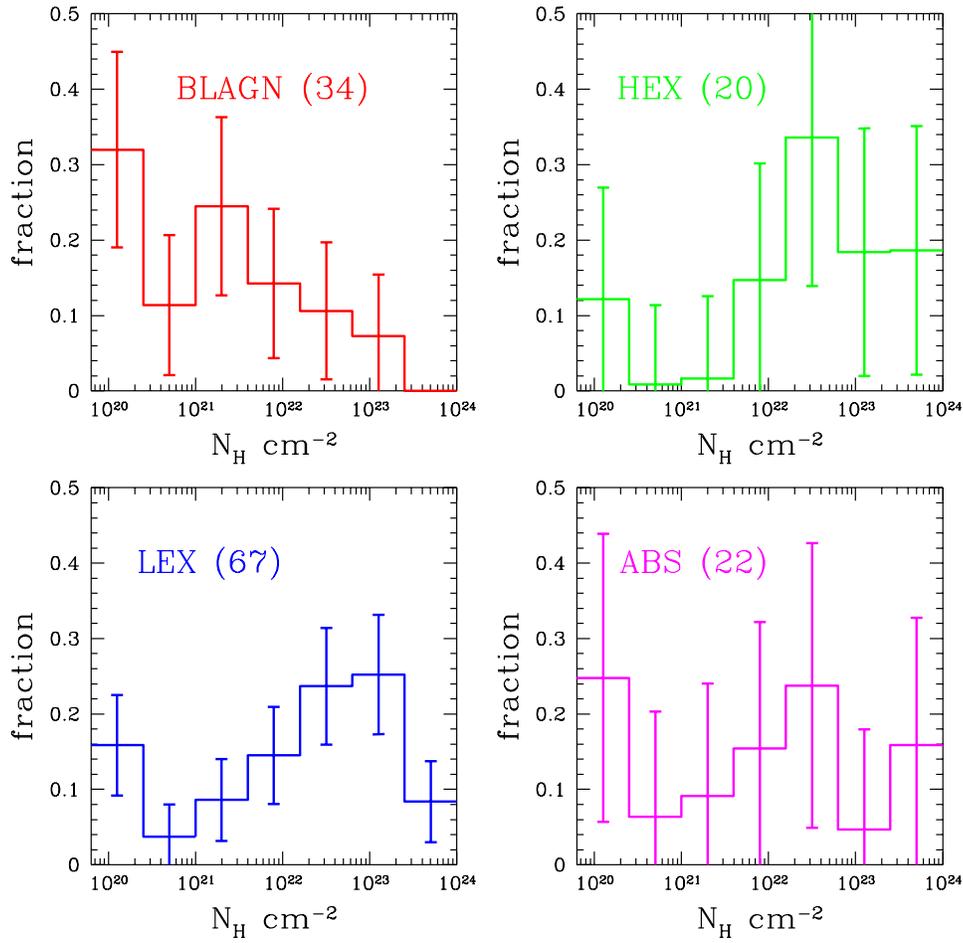}}
\caption{Normalized $N_H$ distribution for the four different optical
classes (see text).
\label{opt_nh}}
\end{figure}

\begin{figure}
\resizebox{\hsize}{!}{\includegraphics{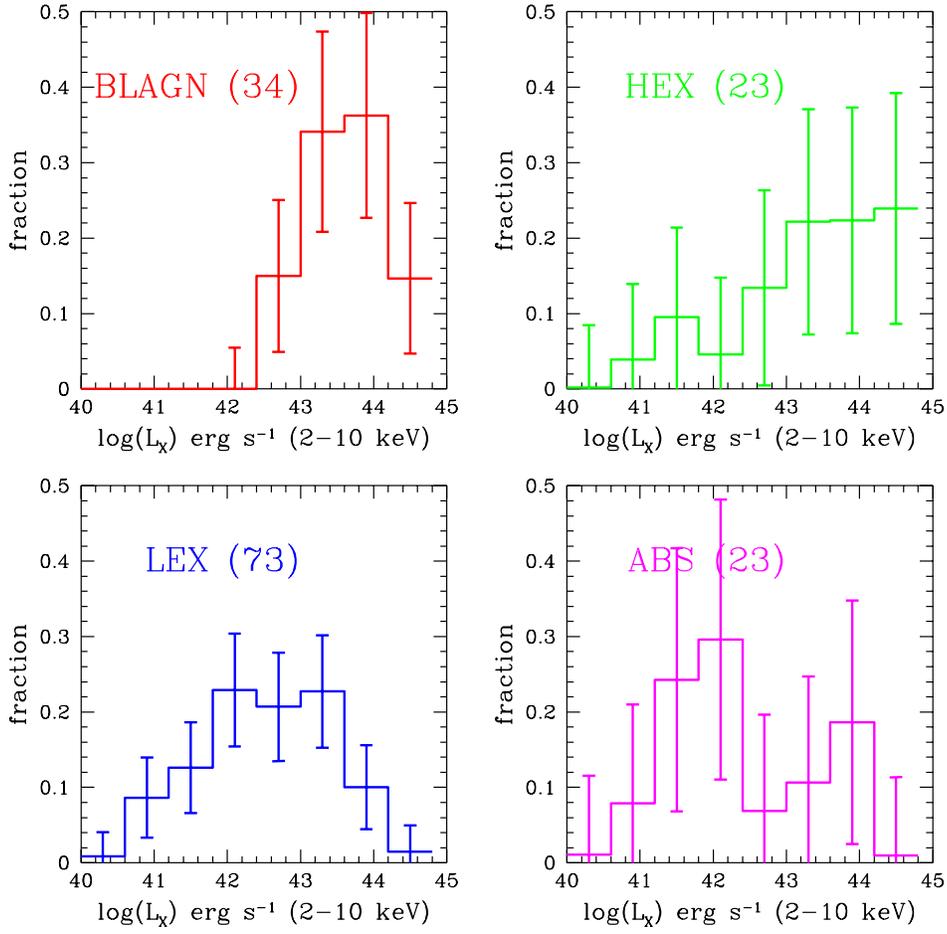}}
\caption{Normalized distribution of the intrinsic rest--frame
luminosity in the hard band for the four different optical classes
(see text).
\label{opt_lh}}
\end{figure}

We also plot the normalized distribution of the intrinsic absorption
and hard luminosities for the four optical classes in Figures
\ref{opt_nh} and \ref{opt_lh}.  Here we account for the statistical
errors by resampling each value according to its error bars, but we do
not introduce any correction for selection effects, since here we are
dominated by optical selection criteria.  We find that, as expected,
the BLAGN class mostly includes AGN with low absorbing column
densities: among the 34 BLAGN, only 7 sources have $N_H> 10^{22}$
cm$^{-2}$; they give a fraction of 0.18 of BLAGN with $N_H> 10^{22}$
cm$^{-2}$, after accounting for statistical errors.  This fraction is
somewhat larger than that found in shallower surveys by Perola et
al. (2004) and in the ChaMP survey by Silverman et al. (2005).
However, we notice that most of the absorbed BLAGN are at $z \geq 2$.
Due to the large errors expected when measuring $N_H$ in high redshift
sources, we do expect a scatter towards high values increasing with
redshift.  A spurious trend $N_H \propto (1+z)^3$ may be visible if we
simply plot the best fit values for $N_H$.  We carefully checked with
simulations with XSPEC that the error bars keep track of this effect,
being larger at higher z.  In these simulations, described in Appendix
C, we show that in the hypothesis of $N_H\simeq 0$ for all the BLAGN
sources, we should expect none of them to have $N_H>0$ at 2 $\sigma $
c.l.  Instead, we find five of them to have $N_H>10^{22}$ cm$^{-2}$ at
2 $\sigma$.  Using the better count statistics and the larger energy
range ($E>0.2$ keV) of XMM (see Streblyanska et al. 2004), the
spectral analysis of 5 of these sources gives absorption in the range
$10^{21}<N_H<10^{22}$ cm$^{-2}$, confirming that these BLAGN have a
non--negligible absorption, but that the Chandra best--fit values are
somewhat higher, possibly due to the limited energy range used which
may hamper the measure of low column densities at high z.
To summarize, we put a strict upper limit of 18\% for absorbed sources
($N_H>10^{22}$ cm$^{-2}$) within BLAGN.

Absorbed AGNs with $N_H> 10^{22}$ cm$^{-2}$ are found mostly in the
HEX and LEX classes (80\% and 60\% respectively).  They are also found
in the ABS class, where, however few sources have $N_H> 10^{22}$
cm$^{-2}$.  We find less evidence for Narrow Line AGN (here classified
as HEX) with low absorption.  We observe only about $\sim 10$\% of
such sources, for which the most likely scenario is severe dilution of
the AGN optical emission by the underlying galaxy.  Therefore, the
simple identification scheme of unabsorbed AGN with optical Type I
(BLAGN) and absorbed AGN with optical Type II (HEX and LEX) is roughly
correct, with uncertainties of less than 20\%.

\begin{figure}
\resizebox{\hsize}{!}{\includegraphics{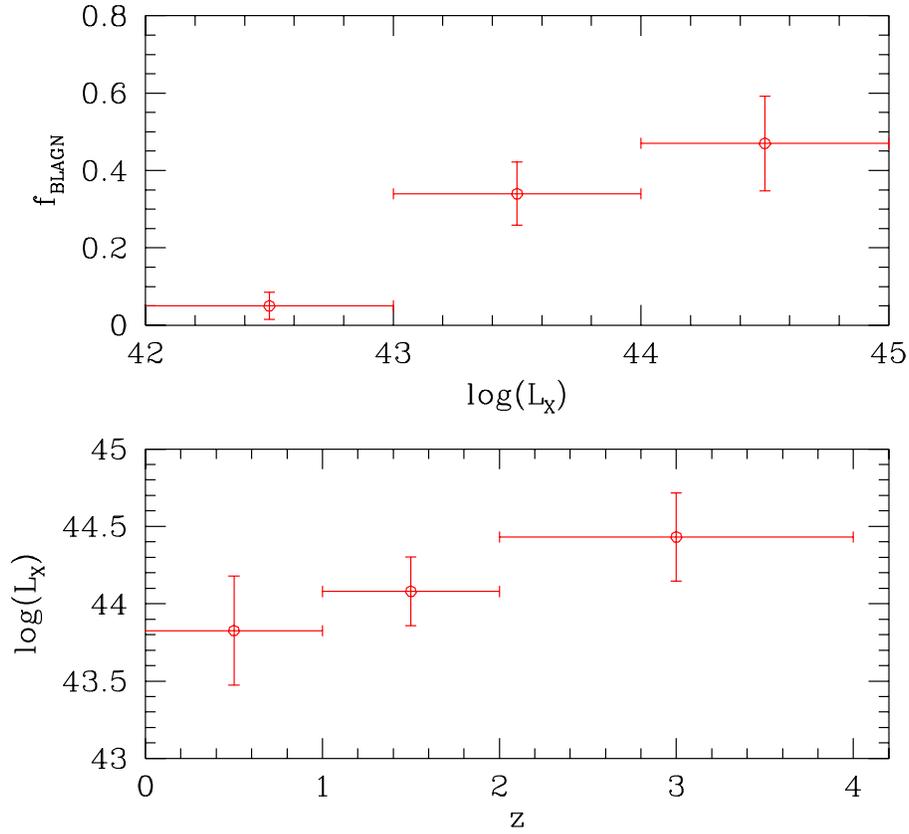}}
\caption{Upper panel: fraction of BLAGN among the sample of sources
with optical redshift as a function of the 0.5-10 keV luminosity bin.
Lower panel: average 0.5-10 keV luminosity of BLAGN as a function of
the redshift bin.  Only sources with $L_X>10^{43}$ erg s$^{-1}$ are
considered in the lower panel.
\label{steffen}}
\end{figure}

As for the hard luminosities (Figures \ref{opt_lh}), we show that the
BLAGN and HEX classes have X--ray luminosities in the range
$10^{42}$--$10^{44}$ erg s$^{-1}$ typical of AGN, with very few
sources below $10^{42}$ erg s$^{-1}$.  The value $10^{42}$ erg
s$^{-1}$ can be considered as an effective luminosity threshold
dividing AGN and normal or star forming galaxies, except for few cases
of galaxies with a strong starburst, which can reach $L_X \sim
10^{42}$ ergs s$^{-1}$ for a star formation rate of about 100
$M_\odot/yr$ (Ranalli et al. 2003).  This luminosity range, where the
presence of normal galaxies is expected to be significant, starts to
be progressively populated in the LEX and ABS classes.  However, also
for the HEX class the majority of the sources have luminosities
$L_X>10^{42}$ erg s$^{-1}$, and only the ABS class is consistent with
being an equal mix of galaxies and AGN.  The distribution of the
intrinsic rest--frame luminosities in the hard bands shows that broad
line AGN have larger intrinsic luminosities than narrow line AGN, as
noted by Barger et al. (2005). In particular, the fraction of BLAGN in
our sample among the sources with optical spectra, is strongly
increasing with luminosity, while their average luminosity is
increasing with redshift, in agreement with the findings of Steffen et
al. (2003), as shown in Figure \ref{steffen}.  
However, due to our small sampling volume at low redshift, to the low
optical spectral completeness of our sample ($\simeq 1/3$), and,
finally, to the possible effect of the stellar dilution that may
hinder the presence of broad lines (see, e.g., Moran et al. 2002) we
do not draw strong conclusion on this aspect.

We note also that, given the intrinsic luminosities and the intrinsic
absorption values found in the remaining subsample of 178 sources
without a clear optical classifications, about 90\% of them are
expected to be secure AGN.  Overall, we find that at least 80\% of the
AGN with spectral ID in our sample agrees with simple AGN unification
models (Antonucci 1993), confirming findings of wider and shallower
surveys (see, e.g., Silverman et al. 2005).

\section{Conclusions}

We presented the detailed spectral analysis of 321 sources in the
CDFS, taking advantage of spectroscopic and photometric redshifts.  We
fitted the source X--ray spectra assuming a default model consisting
in a single power law with intrinsic redshifted absorption (plus a
local absorption frozen to the Galactic value in the direction of the
CDFS) and a Gaussian line at the redshifted energy of the Fe K line
complex.  We look for sources with a spectrum dominated by a
reflection component (Compton--thick candidates) and for sources
showing an unabsorbed scattered component at soft energies.  We are
able to derive the spectral slope distribution for the 82 brightest
sources in the sample and intrinsic absorbing column density for the
whole sample.  Then, from the observed $N_H$ distribution, we derive
the intrinsic $N_H$ distribution for the whole AGN population, after
correcting for incompleteness and for the differential sampling of the
AGN population as a function of intrinsic luminosity and $N_H$
(modelling the luminosity function of AGN after Ueda et al. 2003).  We
accounted for statistical errors in our measures by convolving the
distributions according to the error bars associated to each
measurement.  We also look for evolution in the fraction of absorbed
sources as a function of the redshift.  Our main results are
summarized as follows:

\begin{itemize}
\item We investigate the spectral slope of the intrinsic spectrum for
the 82 sources of the X--ray bright sample, excluding the two
brightest that otherwise would dominate the statistics.  We find that
the average value for the slope of the power law is $\langle \Gamma
\rangle \simeq 1.75 \pm 0.02$, with an intrinsic dispersion of the
order of $\sigma_{int} \simeq 0.30$. 

\item We find no correlation between the spectral index $\Gamma$ and
the intrinsic absorption column density $N_H$ nor the intrinsic
luminosity.  We do not detect any evolution of the average $\Gamma$
with redshift.

\item We select 14 Compton--thick candidates, for which we can only
assess a lower limit to the intrinsic column density of $N_H> 1.5
\times 10^{24}$ cm$^{-2}$.  Due to their low detectability, the
surface density can be as high as $(200 \pm 50)$ deg$^{-2}$.

\item We find significant evidence (at more than 90\% confidence
level) of a Fe line in 20 sources, most of them (14) for the sources
with spectroscopic redshifts.  We also find unabsorbed soft emission,
fit with a power law model with the same slope as the main power law,
possibly associated with a scattered component, in only 8 sources.

\item The intrinsic $N_H$ distribution is well approximated by a
lognormal distribution centered on $\langle log(N_H) \rangle \simeq
23.1$ and with a dispersion $\sigma = 1.1$.  This distribution differs
from that found by Ueda et al. (2003), which shows a broader peak at
lower values of $N_H$.  Our distribution includes the contribution of
many more absorbed AGN, since we explored the faint X--ray flux range,
where strongly absorbed sources dominate in number.  This shows that
the population of Compton--thick AGN (expected to be observed with the
{\sl Spitzer} satellite) is at least partly accounted for in deep
X--ray surveys when all the X--ray selected sources are included.

\item  We find hints that the fraction of absorbed sources is
increasing with redshift, consistently with XRB synthesis models.

\item We find that the simple unification model, i.e. the one--to--one
correspondence of unabsorbed/absorbed X--ray sources to TypeI
AGN--QSO/TypeII AGN--QSOII, is accurate for at least 80\% of the
sources with spectral identification ($\sim 1/3$ of the total X-ray
sample).

\end{itemize}

We remark that once the ongoing or planned spectroscopic follow--up of
the many Chandra and XMM surveys will be completed, the same kind of
detailed spectral analysis will be performed on a much larger number
of sources.  This will allow one to firmly understand the distribution
of spectral properties among AGN, and to suggest improvements to the
unification model in view of the complex relation between X--ray and
optical types.

\begin{acknowledgements} 
The Authors thank Keith Arnaud for help in the use of XSPEC; Alexei
Vikhlinin and Nico Cappelluti for discussion on the reduction of
Chandra data.  We also thank Marcella Brusa, Andrea Comastri, Fabrizio
Fiore, Fabio La Franca, Maurizio Paolillo, Andrew Ptak and Tahir
Yaqoob for discussion.  P. Tozzi acknowledges support under the ESO
visitor program in Garching and the visitor scholarship program the
John Hopkins University during the completion of this work.  We thank
the entire Chandra Team for the high degree of support.
\end{acknowledgements}

\begin{longtable}{llllllllll}
\caption{\label{results}Best fit parameters for the whole sample of
sources in the CDFS with a measured spectroscopic or photometric
redshift.  Error bars correspond to 1 $\sigma$ c.l.  Luminosities are
computed for a flat $\Lambda=0.7$ cosmology and $H_0 = 70$
km/s/Mpc. ID are from Giacconi et al. (2002).  Quality flags with
$Q\geq 1$ indicate optical spectral quality: $Q=1$ corresponds to
spectra with a single optical line identified; $Q=2$ indicates secure
spectra but uncertain counterpart; $Q=3$ indicates secure redshift.
Photometric redshifts are obtained from different methods described in
Zheng et al. (2004): HyperZ ($Q=0.2$), BPZ ($Q=0.3$), COMBO--17
($Q=0.4$ Wolf et al. 2001; 2003; 2004).  When we have consistent
redshift from more than one method, the corresponding quality flag is
the sum of the single $Q$.  }\\ \hline\hline ID & z & Q & $\Gamma$ &
$N_H/10^{22}$ cm$^{-2}$ & $L_{soft}$ erg s$^{-1}$ & $ L_{hard}$ erg
s$^{-1}$ & $F_{0.5-2}$ cgs & $F_{2-10}$ cgs & X--type\\ \hline\hline

1 & 0.347 & 1.6 & $1.54_{-0.21}^{+0.17} $ & $ 0.52_{-0.14}^{+0.16} $ & $ 7.57\times 10^{41} $ & $ 1.73\times 10^{42 } $ & $  1.20\times 10^{-15  } $&  $ 4.78\times 10^{-15 }$ & C--thin \\ 
2 & 0.730 & 0.4 & $3.04_{-0.23}^{+0.31} $ & $ 1.43_{-0.18}^{+0.27} $ & $ 4.37\times 10^{43} $ & $ 9.54\times 10^{42 } $ & $  2.72\times 10^{-15  } $&  $ 2.19\times 10^{-15 }$ & C--thin \\ 
3 & 0.220 & 0.4 & $1.80 $ & $ 0.91_{-0.21}^{+0.22} $ & $ 2.68\times 10^{41} $ & $ 4.19\times 10^{41 } $ & $  6.65\times 10^{-16  } $&  $ 2.91\times 10^{-15 }$ & C--thin \\ 
4 & 1.260 & 1.0 & $1.68_{-0.08}^{+0.14} $ & $ 0.00^{+0.29} $ & $ 2.36\times 10^{43} $ & $ 4.37\times 10^{43 } $ & $  3.28\times 10^{-15  } $&  $ 6.12\times 10^{-15 }$ & C--thin \\ 
6 & 2.460 & 0.2 & $1.91_{-0.12}^{+0.13} $ & $ 1.80_{-0.95}^{+0.96} $ & $ 2.09\times 10^{44} $ & $ 2.73\times 10^{44 } $ & $  3.79\times 10^{-15  } $&  $ 6.40\times 10^{-15 }$ & C--thin \\
\end{longtable}

\newpage

\begin{longtable}{llllll}
\caption{\label{optype}Comparison of the optical and X--ray classification  }\\
\hline\hline 
 & BLAGN  & HEX  &  LEX  &  ABS & Undet\\
X-ray AGN--1 and QSO--1  &  27    &  2   &   13  &   3  &  40   \\
X--ray AGN--2 and QSO--2 &   7    &  18  &   42  &  10  & 118   \\
X--ray galaxy            &   0    &  0   &   12  &   9  &  21   \\
\hline\hline 
\end{longtable}

\newpage

\newpage

\newpage
\clearpage

\newpage

\newpage

\clearpage
\newpage

\appendix

\section{Fitted X--ray spectra}

The background--subtracted, unfolded spectra of the sources analyzed
in this Paper, along with the best fit models, are shown in Figures
\ref{spectra1}--\ref{spectra22}.  Spectra are binned for display
purpose only, with the simple criterion of having at least 20 counts
or a signal--to--noise of 3 in each bin for sources with more than
$100$ net detected counts.  Weaker sources spectra are binned with at
least 10 counts or a signal--to--noise of 2 in each bin.  We recall
that the binning is used only to plotting purpose, while the unbinned
spectra are used when performing the fit, as described in the text.
Each source is fitted with the best--fit model, which is the canonical
Compton--thin plus gaussian line model for the large majority of the
sources (Figures \ref{spectra1}--\ref{spectra22}).  For 14
Compton--thick candidates we show the reflection model plus the
gaussian line (Figures \ref{spectra20}), while in 8 cases we add a
soft component (Figures \ref{spectra21}).  We remind that a reversed
edge at $2.07$ keV is added to each spectrum to take into account a
small increase in the efficiency of the ACIS detectors which is not
yet included in the ACIS response functions in CALDB2.26.  Such a
feature, visible as a small step just above 2 keV, should not be
considered intrinsic to the sources.

\newpage
\clearpage

\begin{figure} 

\vskip 1cm
\begin{center}
{\bf for this figure see http://adlibitum.oat.ts.astro.it/tozzi/spectral\_CDFS/paper\_1Mspec.ps}
\end{center}
\vskip 1cm






\caption{Unfolded spectra and best fit model for the 299 C--thin
sources (fitted with a {\tt zwabs pow} model with $\Gamma = 1.8$ plus
a narrow Gaussian line at 6.4 keV rest frame). }
\label{spectra1} 
\end{figure} 


\begin{figure}
\vskip 1cm
\begin{center}
{\bf for this figure  see http://adlibitum.oat.ts.astro.it/tozzi/spectral\_CDFS/paper\_1Mspec.ps}
\end{center}
\vskip 1cm






\caption{Figure \ref{spectra1}, continued. }
\label{spectra2} 
\end{figure} 


\begin{figure}

\vskip 1cm
\begin{center}
{\bf for this figure  see http://adlibitum.oat.ts.astro.it/tozzi/spectral\_CDFS/paper\_1Mspec.ps}
\end{center}
\vskip 1cm



%


\caption{Figure \ref{spectra1}, continued. }
\label{spectra3} 
\end{figure} 


\begin{figure}

\vskip 1cm
\begin{center}
{\bf for this figure  see http://adlibitum.oat.ts.astro.it/tozzi/spectral\_CDFS/paper\_1Mspec.ps}
\end{center}
\vskip 1cm






\caption{Figure \ref{spectra1}, continued. }
\label{spectra4} 
\end{figure} 


\begin{figure} 

\vskip 1cm
\begin{center}
{\bf for this figure  see http://adlibitum.oat.ts.astro.it/tozzi/spectral\_CDFS/paper\_1Mspec.ps}
\end{center}
\vskip 1cm






\caption{Figure \ref{spectra1}, continued. }
\label{spectra5} 
\end{figure} 


\begin{figure} 

\vskip 1cm
\begin{center}
{\bf for this figure  see http://adlibitum.oat.ts.astro.it/tozzi/spectral\_CDFS/paper\_1Mspec.ps}
\end{center}
\vskip 1cm



%
%

\caption{Figure \ref{spectra1}, continued. }
\label{spectra6} 
\end{figure} 


\begin{figure} 

\vskip 1cm
\begin{center}
{\bf for this figure  see http://adlibitum.oat.ts.astro.it/tozzi/spectral\_CDFS/paper\_1Mspec.ps}
\end{center}
\vskip 1cm






\caption{Figure \ref{spectra1}, continued. }
\label{spectra7} 
\end{figure} 


\begin{figure} 

\vskip 1cm
\begin{center}
{\bf for this figure  see http://adlibitum.oat.ts.astro.it/tozzi/spectral\_CDFS/paper\_1Mspec.ps}
\end{center}
\vskip 1cm






\caption{Figure \ref{spectra1}, continued. }
\label{spectra8} 
\end{figure} 


\begin{figure} 

\vskip 1cm
\begin{center}
{\bf for this figure  see http://adlibitum.oat.ts.astro.it/tozzi/spectral\_CDFS/paper\_1Mspec.ps}
\end{center}
\vskip 1cm






\caption{Figure \ref{spectra1}, continued. }
\label{spectra9} 

\end{figure} 


\begin{figure}

\vskip 1cm
\begin{center}
{\bf for this figure  see http://adlibitum.oat.ts.astro.it/tozzi/spectral\_CDFS/paper\_1Mspec.ps}
\end{center}
\vskip 1cm
%





\caption{Figure \ref{spectra1}, continued. }
\label{spectra10} 
\end{figure} 

\clearpage

\begin{figure} 

\vskip 1cm
\begin{center}
{\bf for this figure  see http://adlibitum.oat.ts.astro.it/tozzi/spectral\_CDFS/paper\_1Mspec.ps}
\end{center}
\vskip 1cm






\caption{Figure \ref{spectra1}, continued. }
\label{spectra11} 

\end{figure}


\begin{figure} 

\vskip 1cm
\begin{center}
{\bf for this figure  see http://adlibitum.oat.ts.astro.it/tozzi/spectral\_CDFS/paper\_1Mspec.ps}
\end{center}
\vskip 1cm




%

\caption{Figure \ref{spectra1}, continued.}
\label{spectra12} 

\end{figure} 


\begin{figure} 

\vskip 1cm
\begin{center}
{\bf for this figure  see http://adlibitum.oat.ts.astro.it/tozzi/spectral\_CDFS/paper\_1Mspec.ps}
\end{center}
\vskip 1cm






\caption{Figure \ref{spectra1}, continued. }
\label{spectra13} 

\end{figure} 


\begin{figure} 

\vskip 1cm
\begin{center}
{\bf for this figure  see http://adlibitum.oat.ts.astro.it/tozzi/spectral\_CDFS/paper\_1Mspec.ps}
\end{center}
\vskip 1cm






\caption{Figure \ref{spectra1}, continued. }
\label{spectra14} 

\end{figure} 


\begin{figure} 

\vskip 1cm
\begin{center}
{\bf for this figure  see http://adlibitum.oat.ts.astro.it/tozzi/spectral\_CDFS/paper\_1Mspec.ps}
\end{center}
\vskip 1cm






\caption{Figure \ref{spectra1}, continued. }
\label{spectra15} 

\end{figure} 


\begin{figure} 

\vskip 1cm
\begin{center}
{\bf for this figure  see http://adlibitum.oat.ts.astro.it/tozzi/spectral\_CDFS/paper\_1Mspec.ps}
\end{center}
\vskip 1cm






\caption{Figure \ref{spectra1}, continued. }
\label{spectra16} 
\end{figure}


\begin{figure} 

\vskip 1cm
\begin{center}
{\bf for this figure  see http://adlibitum.oat.ts.astro.it/tozzi/spectral\_CDFS/paper\_1Mspec.ps}
\end{center}
\vskip 1cm






\caption{Figure \ref{spectra1}, continued. }
\label{spectra17} 

\end{figure} 


\begin{figure} 

\vskip 1cm
\begin{center}
{\bf for this figure  see http://adlibitum.oat.ts.astro.it/tozzi/spectral\_CDFS/paper\_1Mspec.ps}
\end{center}
\vskip 1cm




%

\caption{Figure \ref{spectra1}, continued. }
\label{spectra18} 
\end{figure} 


\begin{figure} 

\vskip 1cm
\begin{center}
{\bf for this figure  see http://adlibitum.oat.ts.astro.it/tozzi/spectral\_CDFS/paper\_1Mspec.ps}
\end{center}
\vskip 1cm






\caption{Figure \ref{spectra1}, continued. }
\label{spectra19} 

\end{figure} 


\begin{figure} 

\vskip 1cm
\begin{center}
{\bf for this figure  see http://adlibitum.oat.ts.astro.it/tozzi/spectral\_CDFS/paper\_1Mspec.ps}
\end{center}
\vskip 1cm






\caption{Figure \ref{spectra1}, continued. } 
\label{spectra20}
\end{figure} 


\begin{figure} 

\vskip 1cm
\begin{center}
{\bf for this figure  see http://adlibitum.oat.ts.astro.it/tozzi/spectral\_CDFS/paper\_1Mspec.ps}
\end{center}
\vskip 1cm






\caption{Unfolded spectra and best fit model for the 14 C--thick
sources (fitted with a {\tt pexrav} model). }
\label{spectra21}
\end{figure} 


\begin{figure} 

\vskip 1cm
\begin{center}
{\bf for this figure see http://adlibitum.oat.ts.astro.it/tozzi/spectral\_CDFS/paper\_1Mspec.ps}
\end{center}
\vskip 1cm




\caption{Unfolded spectra and best fit model for the 8 Soft--C
sources (fitted with a {\tt pow + zwabs pow} model). }
\label{spectra22}
\end{figure} 

\newpage

\section{Selection of Compton--thick candidates: spectral
 simulations}

We describe here the strategy we adopted in order to select
Compton--thick candidates on the basis of the X--ray spectrum.  We
also want to evaluate the efficiency of our method, and keep control
on the fraction of spurious candidates.  First, we select a subsample
of 110 sources choosen among the 321 sources of the sample because of
their flat spectrum, with best--fit slope $\Gamma \leq 1$ when fitted
with a simple power law without absorption.  This subsample is
expected to include the most obscured component of the XRB (see
Civano, Comastri \& Brusa 2005).  Therefore we assume that all the
Compton thick sources are included in this subsample.

\begin{figure}
 \resizebox{\hsize}{!}{\includegraphics{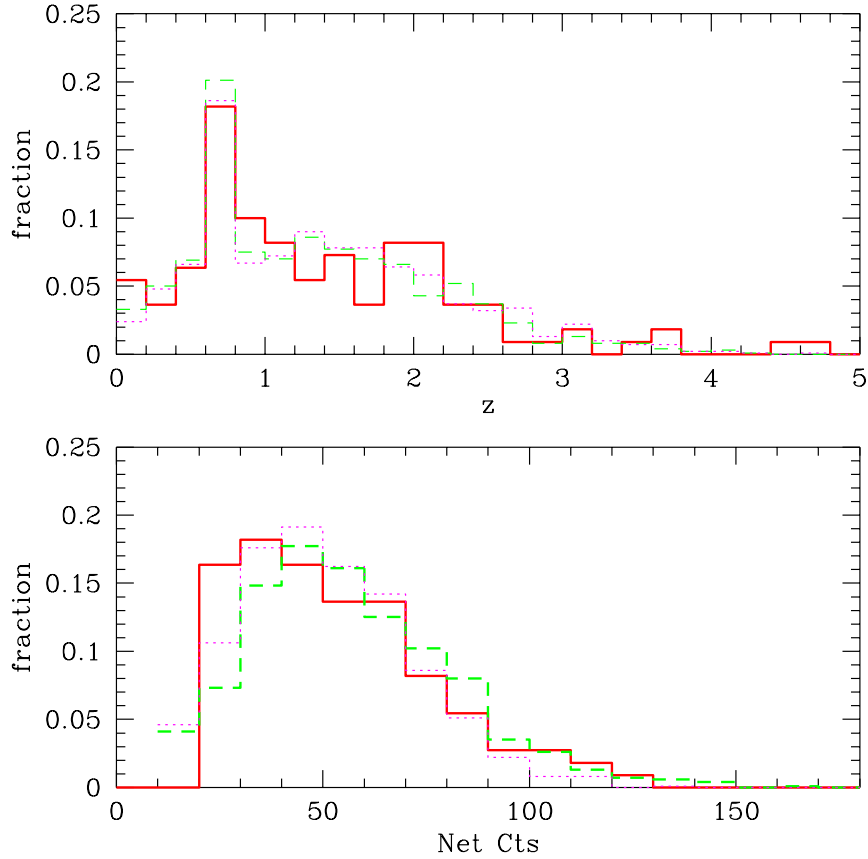}}
 \caption{Normalized redshift distribution (upper panel) and net
 detected counts (lower panel) for the subsample of real sources with
 flat spectrum (continuous lines) and for sources simulated with a
 {\tt pexrav} spectral model (dashed lines -- Simulation 1) and with a
 {\tt zwabs pow} spectral model (dotted lines -- Simulation 2).
 \label{sim_dist}}
\end{figure}

We also assume for simplicity, that all the sources can be described
by two possible spectral shape: an absorbed power law for Compton thin
sources, and a pure reflection for Compton thick sources.  With the
command {\tt fakeit} within XSPEC, we simulated 1000 sources with a
pure cold reflection spectrum, {\tt pexrav} in XSPEC, with $\Gamma$
fixed to 1.8 and all the other parameters set to the default values
(Simulation 1).  Each simulated source is assigned a redshift and a
normalization according to the distribution of the redshifts and the
net detected counts of the subsample of real sources.  Then, we
simulated another 1000 sources with an absorbed power law ({\tt zwabs
pow} model), with a similar redshift and net detected counts
distributions (Simulation 2).  In Figure \ref{sim_dist} we compare the
redshift and net detected counts distributions of the simulated
sources with that of the parent sample of real sources.  The values of
$N_H$ for the sources simulated with the model {\tt zwabs pow}, are
consistently extracted from the distribution we found in the paper
(but only for $N_H < 10^{24}$ cm$^{-2}$ to exclude Compton thick
sources).

\begin{figure}
 \resizebox{\hsize}{!}{\includegraphics{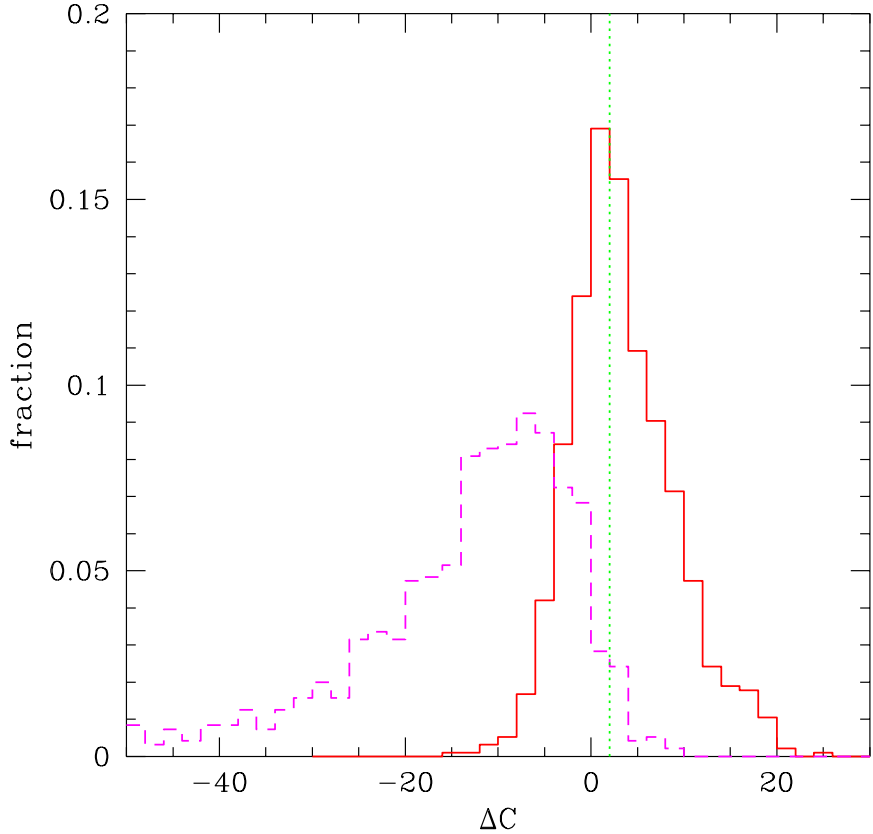}}
 \caption{Normalized distribution of $\Delta C$ for Simulation 1
 (continuous line) and for Simulation 2 (dashed line).  The vertical
 dotted line correspond to the choosen threshold $\Delta C = 2$.
  \label{deltac}}  
\end{figure}

Then, we analyzed the two sets of simulation both with the {\tt
pexrav} (appropriate only for Simulation 1) and {\tt zwabs pow}
(appropriate only for Simulation 2) model.  We verified that in the
the first case we succesfully recover the input values for the
normalization of the {\tt pexrav} spectra, and in the second case the
input values for $N_H$, within the errors.  Finally, we compute the
difference between the Cash-statistics obtained with the {\tt zwabs
pow} model and that obtained with the {\tt pexrav} model: $\Delta C =
C_{zwapow}-C_{pexrav}$.  The normalized distributions of the values of
$\Delta C$ are shown in Figure \ref{deltac}.  We recall that we have
two free parameters for the {\tt zwabs pow} model and only one for the
{\tt pexrav} model; this explains why the distribution of Simulation 2
has a much larger tail at negative values of $\Delta C$.  Our goal is
to use these distributions to choose a fixed threshold $\tilde \Delta
C$ that allows us to select Compton thick candidates among the parent
sample.  The optimal choice would minimize the number of Compton thin
sources mistakenly included in the C--thick sample, at the same time
recovering the largest fraction of the Compton thick population.

The distribution of $\Delta C$ for Simulation 1 is skewed towards
large positive values, as expected since the {\tt pexrav} model is the
correct one.  The tail at low values of $\Delta C$ is a measure of how
many Compton thick sources may be missed when choosing a fixed
threshold in $\Delta C$.  The distribution of $\Delta C$ for
Simulation 2 is centered around negative values, since here the {\tt
pexrav} model is not appropriate.  Therefore, the tail at high values
of $\Delta C$ is a measure of how many sources with an actual {\tt
zwabs pow} spectrum are mistakenly selected as C--thick candidates for
a fixed threshold in $\Delta C$.

The simplifying assumption that our subsample of real sources includes
only C--thin and C--thick sources, reads $N_{tot} =
N_{C-thick}+N_{C-thin}$ where $N_{tot}$ is the total number of sources
in the subsample (here 110).  We collect $N_C$ sources as C--thick
candidates by selecting the sources from the parent sample for which
$\Delta C > \tilde \Delta C$.  The expected value for $N_C$ is:

\begin{equation}
N_C = N_{C-thick}\, * \, F1 + N_{C-thin}\, *\, F2 \, ,
\end{equation}
\noindent

where $F_1$ and $F_2$ are the probabilities that a C--thick source is
correctly recovered, and that a C--thin source is mistakenly included
among the C--thick candidates, respectively.  We can estimate $F_1$
and $F_2$ by integrating the two distributions for $\Delta C > \tilde
\Delta C$.  

Therefore, the actual fraction of the C--thick sources in the parent
sample $f_{CT} \equiv N_{C-thick}/N_{tot}$ can be estimated as:

\begin{equation}
f_{CT} = \Large( N_C / N_{tot} - F_2\Large) /( F_1-F_2).
\label{fct_eq}
\end{equation}
\noindent

\begin{figure}
\resizebox{\hsize}{!}{\includegraphics{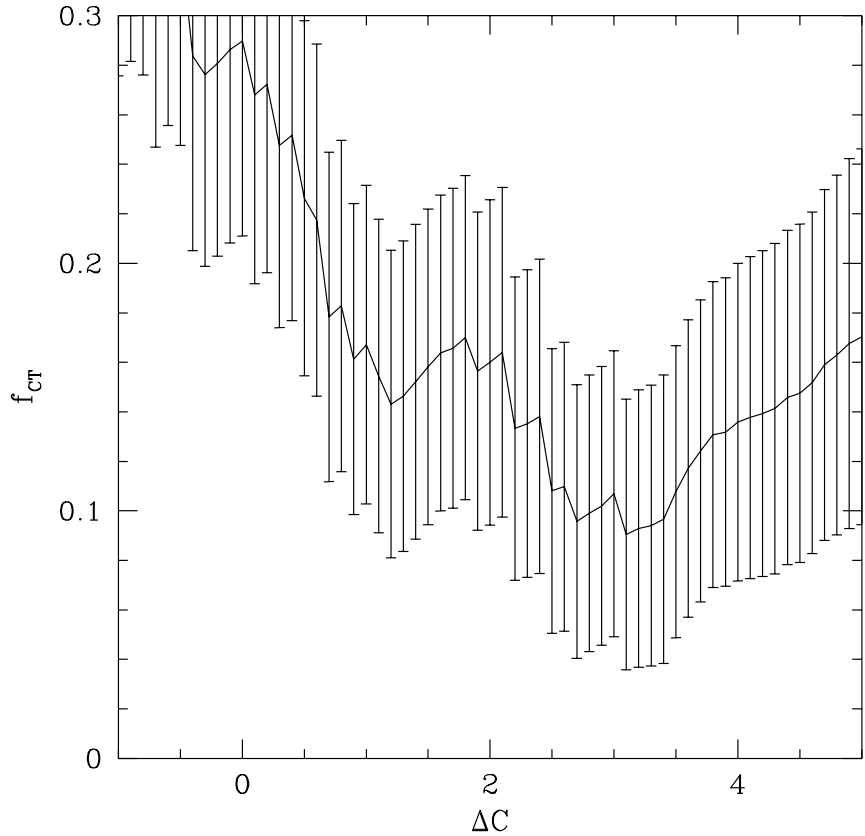}}
 \caption{Estimated value of the fraction of C--thick sources among
 the parent sample (110 sources) as a function of $\Delta C$ according
 to eq. \ref{fct_eq}.
\label{fct}}  
\end{figure}

Our estimated $f_{CT}$ should not depend on $\tilde \Delta C$ if our
initial assumption $N_{tot} = N_{C-thick}+N_{C-thin}$ is correct.
However, we know that the picture may be complicated by the presence
of sources with soft component, or the lack of a proper treatment of
the Compton scattering when $N_H$ approach the Compton thick value of
$1.5 \times 10^{24}$ cm$^{-2}$ (which would require the use of the
model {\tt plcabs}).  However, a more detailed treatmend would go
beyond the scope of this Paper.  We find that for $\Delta C > 1$ the
expected values for $f_{CT}$ ranges between 0.10 and 0.20 (see Figure
\ref{fct}).  Note here that this fraction is computed among the parent
sample of 110 sources, therefore it corresponds to a number between
only 10 and 20 C--thick sources in the whole CDFS sample.  

\begin{figure}
\resizebox{\hsize}{!}{\includegraphics{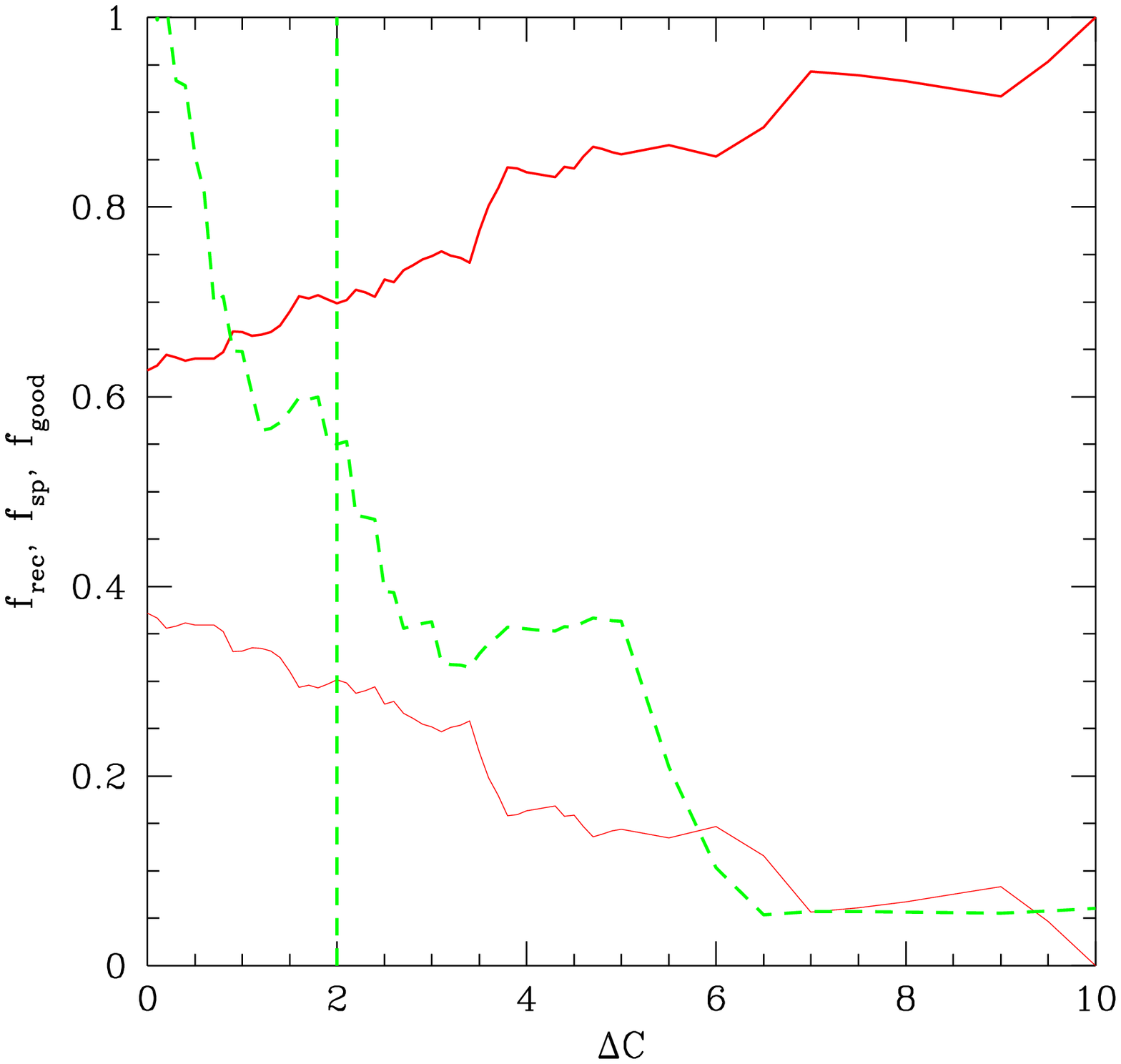}}
\caption{Upper solid line: expected fraction of true C--thick sources
among the candidates ($f_{good}$) versus $\Delta C$; lower solid line:
expected fraction of spurious sources among the candidates ($f_{sp}$);
dashed line: fraction of the total C--thick source population actually
recovered ($f_{rec}$).
\label{comp}}  
\end{figure}

Obviously, for higher values of $\Delta C$, the quality of the
C--cthick condidates sample is increasing, while the fraction of true
C--thick sources actually recovered drops.  Assuming $f_{CT} \simeq
0.15$, we plot in figure \ref{comp} the following quantities as a
function of $\Delta C$: the expected fraction of true C--thick sources
among the candidates ($f_{good}$); the expected fraction of spurious
sources among the candidates ($f_{sp}$); the fraction of the total
C--thick source population actually recovered ($f_{rec}$).  We notice
that for $\Delta C \leq 2$ more than 50\% of the total C--thick source
population is recovered.  Therefore we assume $\tilde \Delta C = 2$.

The number of C--thick candidates we find for $\tilde \Delta C = 2$ is
14 (see text).  The number of spurious sources among the C--thick
candidates turns out to be still significant, between 3 and 5.  To
summarize, we demonstrated here that a selection of C--thick sources
on the basis of the X--ray spectrum is feasible.  We also show that we
can quantify the completeness and the contamination of our C--thick
candidate sample.  We also notice that the level of contamination is
not negligible, pointing towards the need of a more sophisticated
X--ray spectral analysis, including, for example, the systematic
search for the Fe line expected more frequently in
reflection--dominated spectra, or considering the presence of a soft
component that can mimick a flat spectrum (see Weaver et al. 1996).
Overall, we believe that a refined version of this approach can
constitute a valuable tool to look for Compton thick sources on the
basis of the X--ray data only.

\newpage 

\newpage

\section{Measures of $N_H$ at high--z: simulations}

The best fit values of $N_H$ may show a spurious trend with redshift,
due to the increasing difficulty of measuring $N_H$ when the
rest--frame soft band, which is most sensible to the intrinsic
absorption, is shifted out of the Chandra energy range.  What happens
typically is that the error bars are so large that, while the upper
limits to $N_H$ increase with redshift, the best fit values may
fluctuate at large positive values also when the source has a
negligible intrinsic absorption.  We take into account part of this
effect by resampling the value of $N_H$ according to the error bars,
however since $N_H$ is always defined as positive, it is hard to avoid
an average trend of increasing intrinsic absorption with redshift.  In
particular, we asked ourselves if the absorbed BLAGN found at $z>2$
may be spurious (see \S 7).

\begin{figure} 
\resizebox{\hsize}{!}{\includegraphics{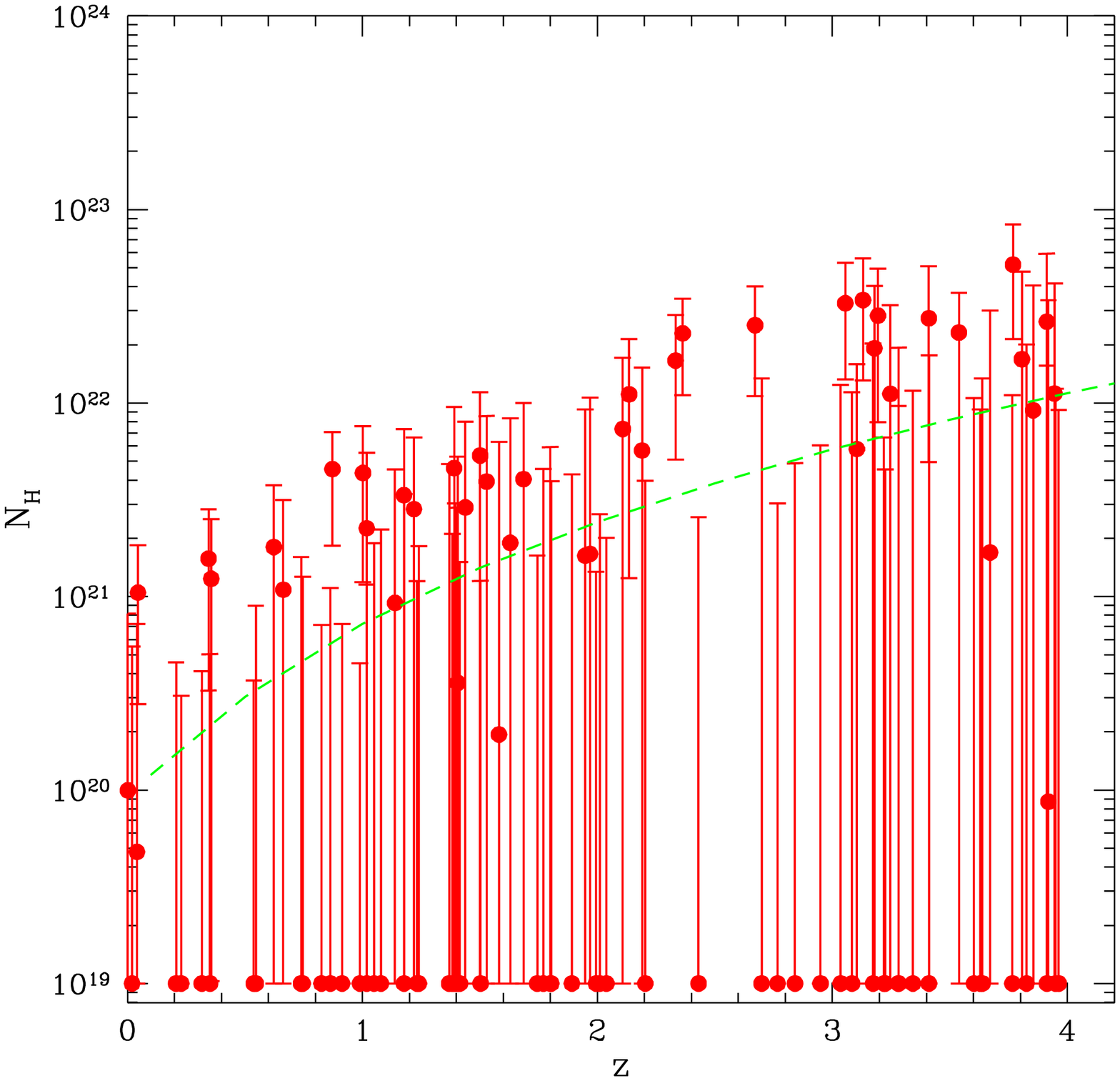}\includegraphics{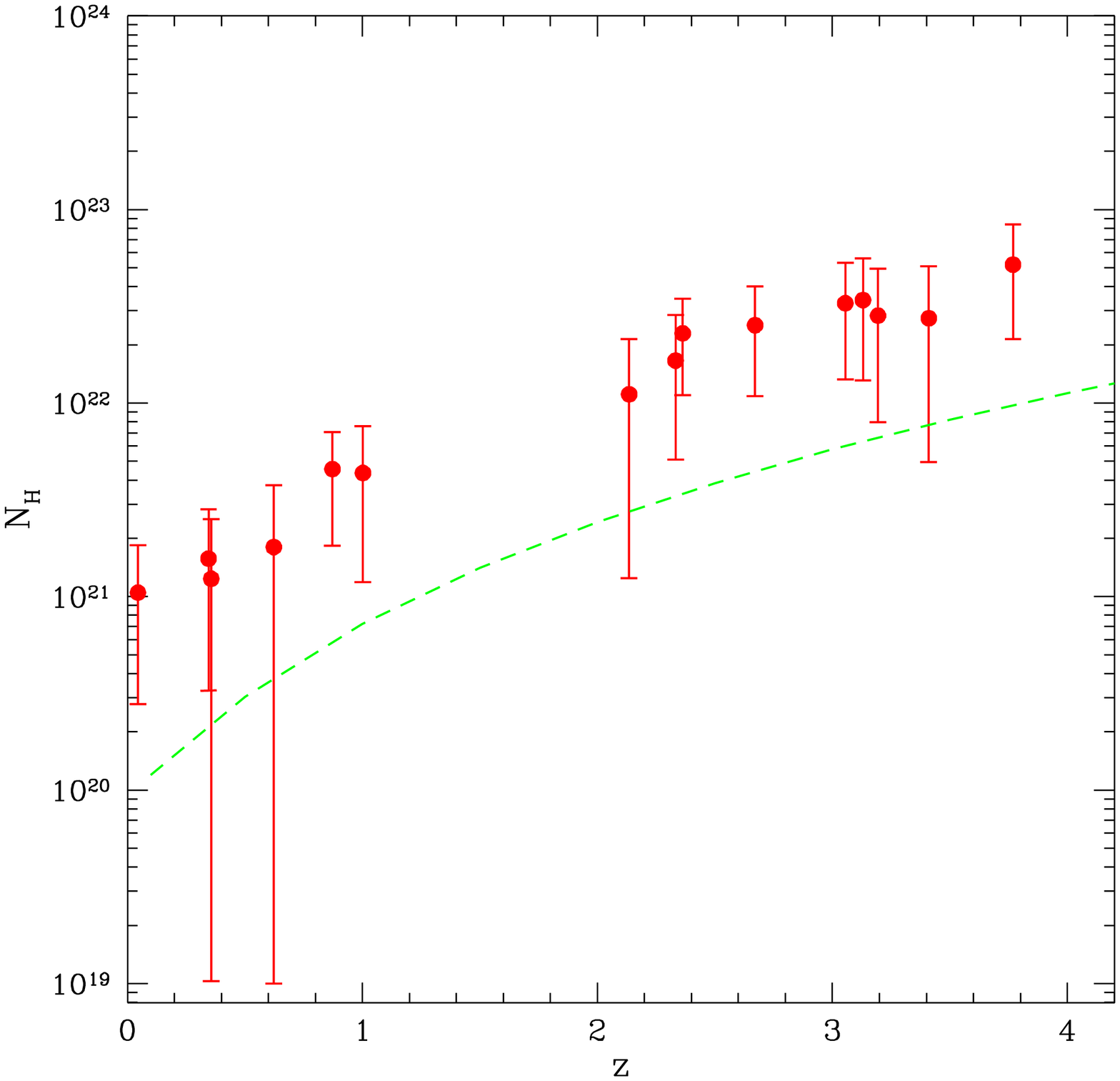}}
\resizebox{\hsize}{!}{\includegraphics{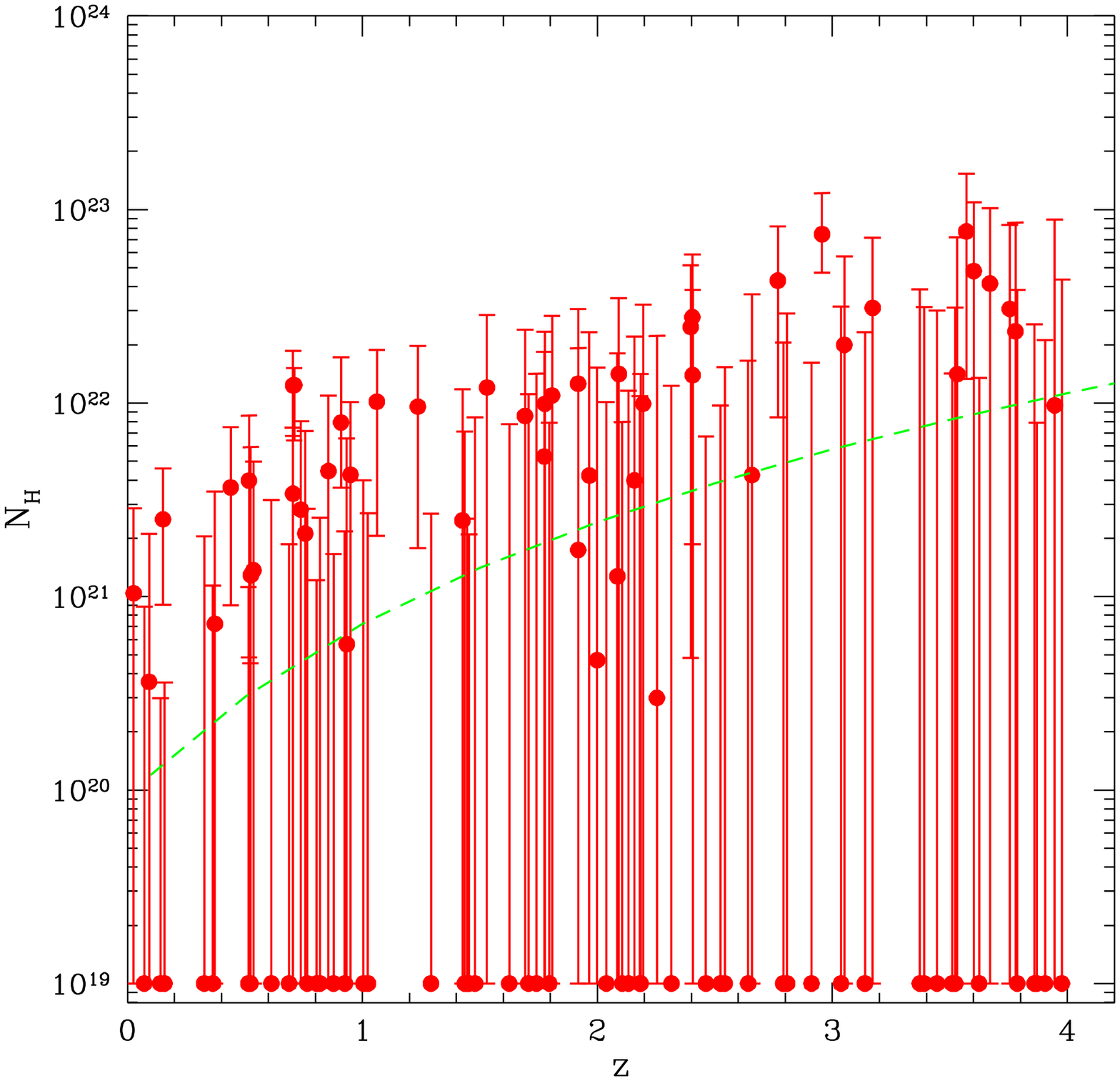}\includegraphics{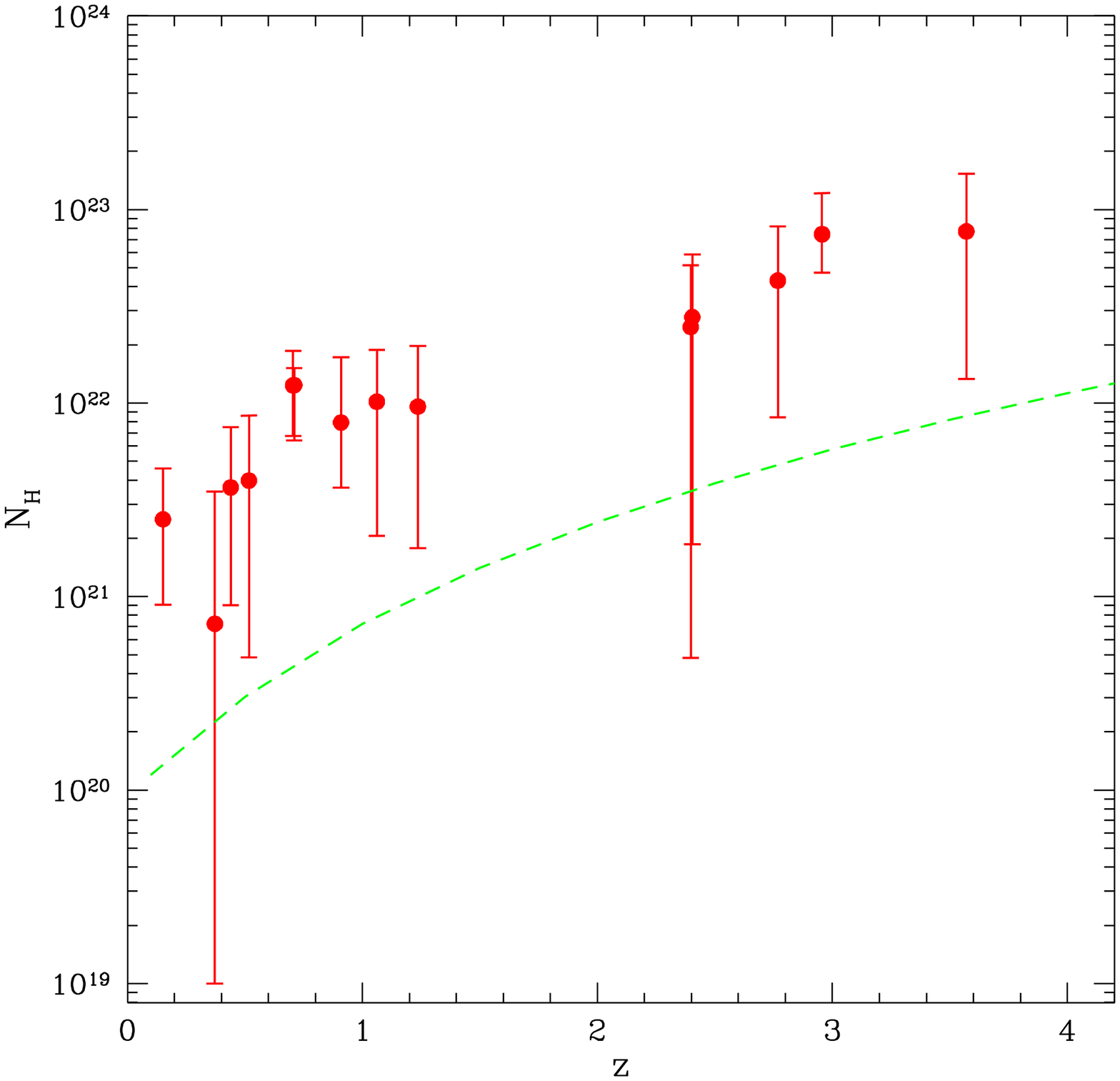}}
\resizebox{\hsize}{!}{\includegraphics{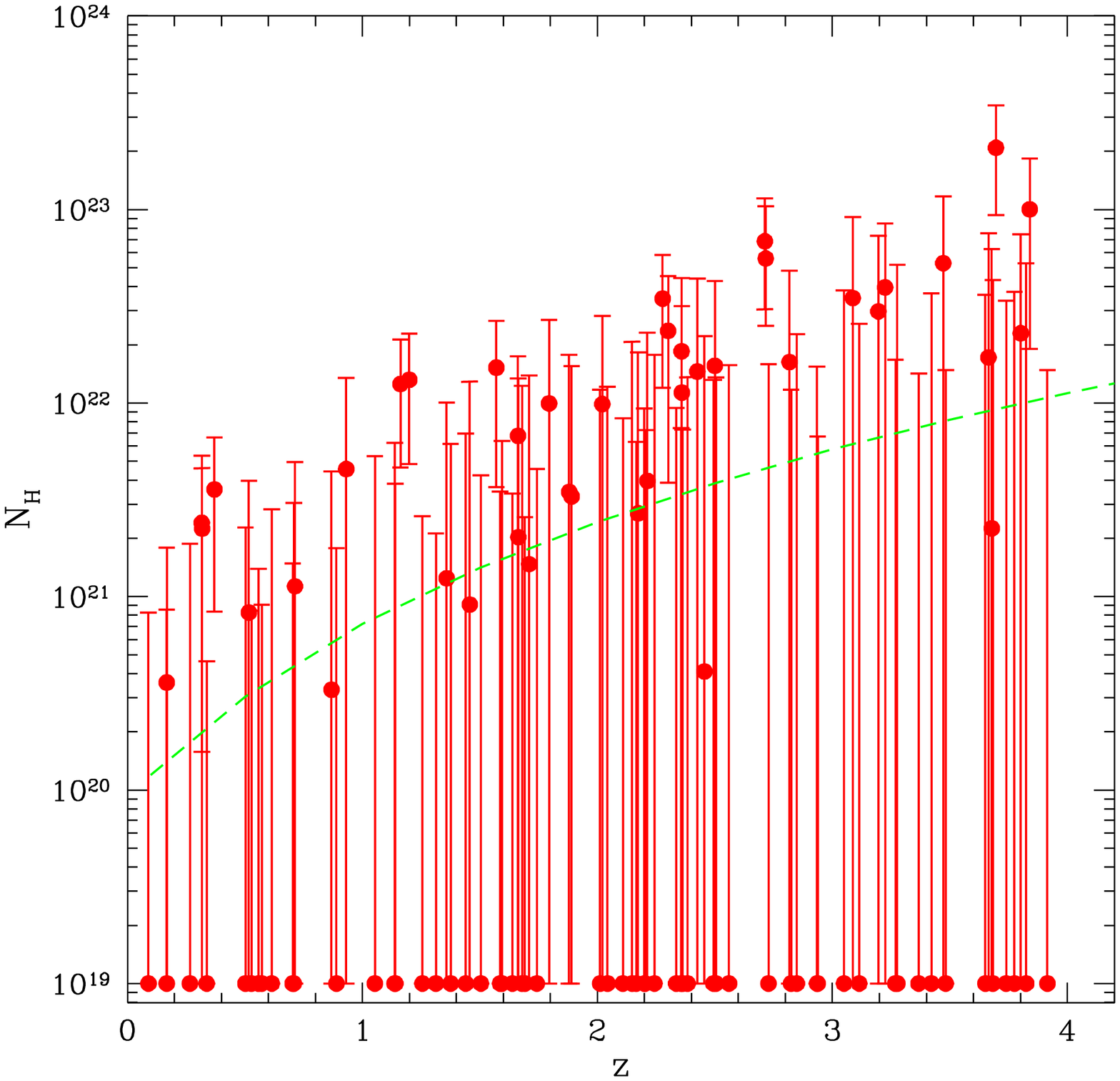}\includegraphics{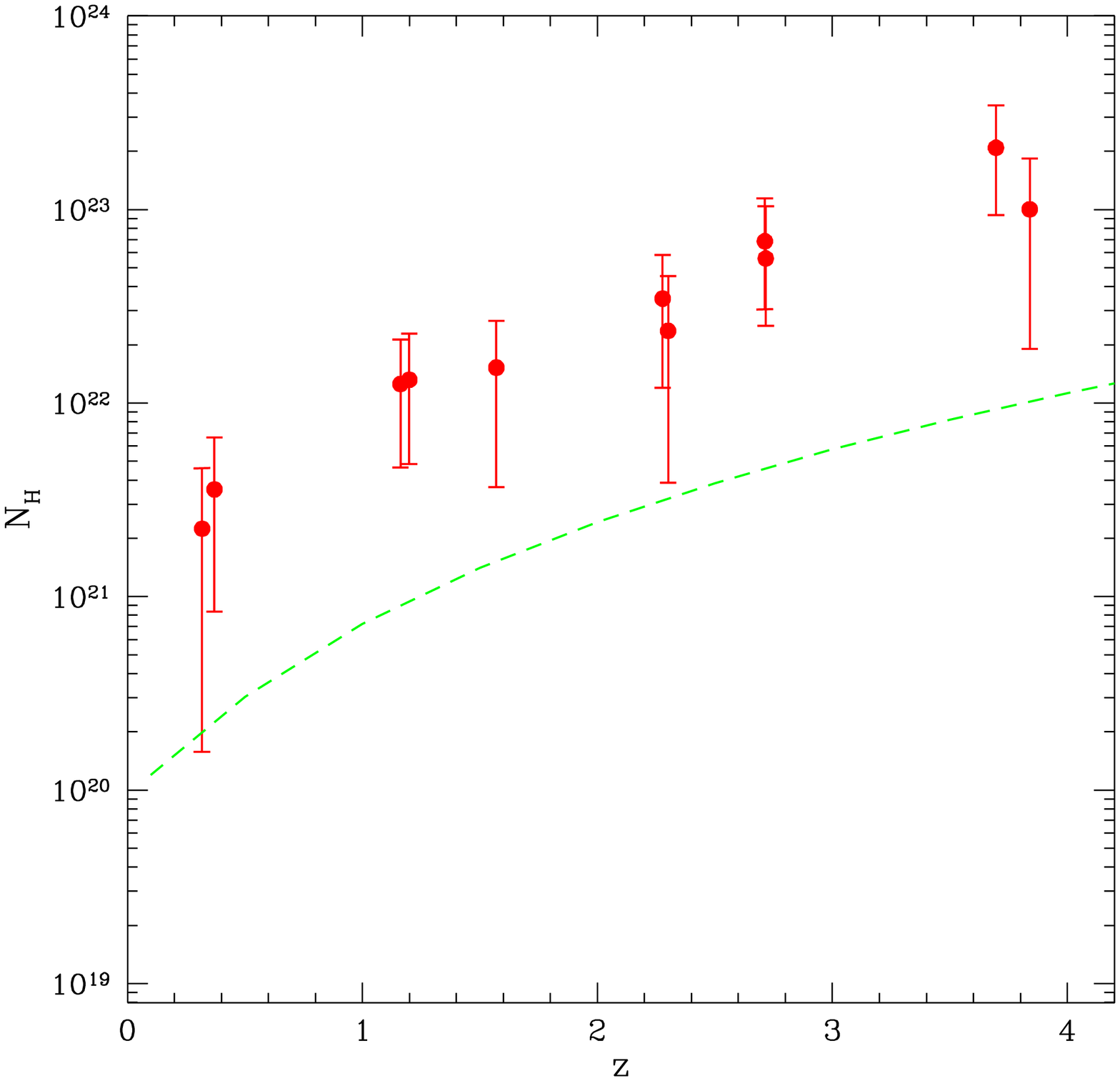}}
\caption{Lef panels: Best fit values of $N_H$ as a function of
redshift for simulated sources with input values $N_H = 9\times
10^{19}$ cm$^{-2}$.  The dashed line is $ 9\times 10^{19} (1+z)^3$.
Error bars correspond to 1 $\sigma$.  The typical net detected counts
are 480, 150 and 80 from top to bottom.  Right panels: values of $N_H$
for which $N_H>0$ at 1 $\sigma$.  }
\label{nh_sim}
\end{figure}

To investigate this effect, we performed three sets of simulations
(100 spectra each) of sources with negligible absorption (equal to the
galactic value $9\times 10^{19}$ cm$^{-2}$) with redshift distributed
uniformly in the range $z=0-4$.  The average number of net detected
counts in each set of simulations is 490, 150 and 80.  In Figure
\ref{nh_sim} we show the best fit values of $N_H$ (with 1 $\sigma$
error bars) plotted versus redshift.  We also plot separately the
cases in which $N_H > 0$ at more than 1 $\sigma$.  We notice that,
while the upper limits increases following approximately the $(1+z)^3$
law, the number of false detections of a non--negligible intrinsic
absorption ($N_H >0$ at more than 1 $\sigma$) is not increasing
significantly with redshift.  In addition, we find that at 2 $\sigma$
confidence level, all the values of $N_H$ are consistent with
negligible absorption.

\begin{figure} 
\resizebox{\hsize}{!}{\includegraphics{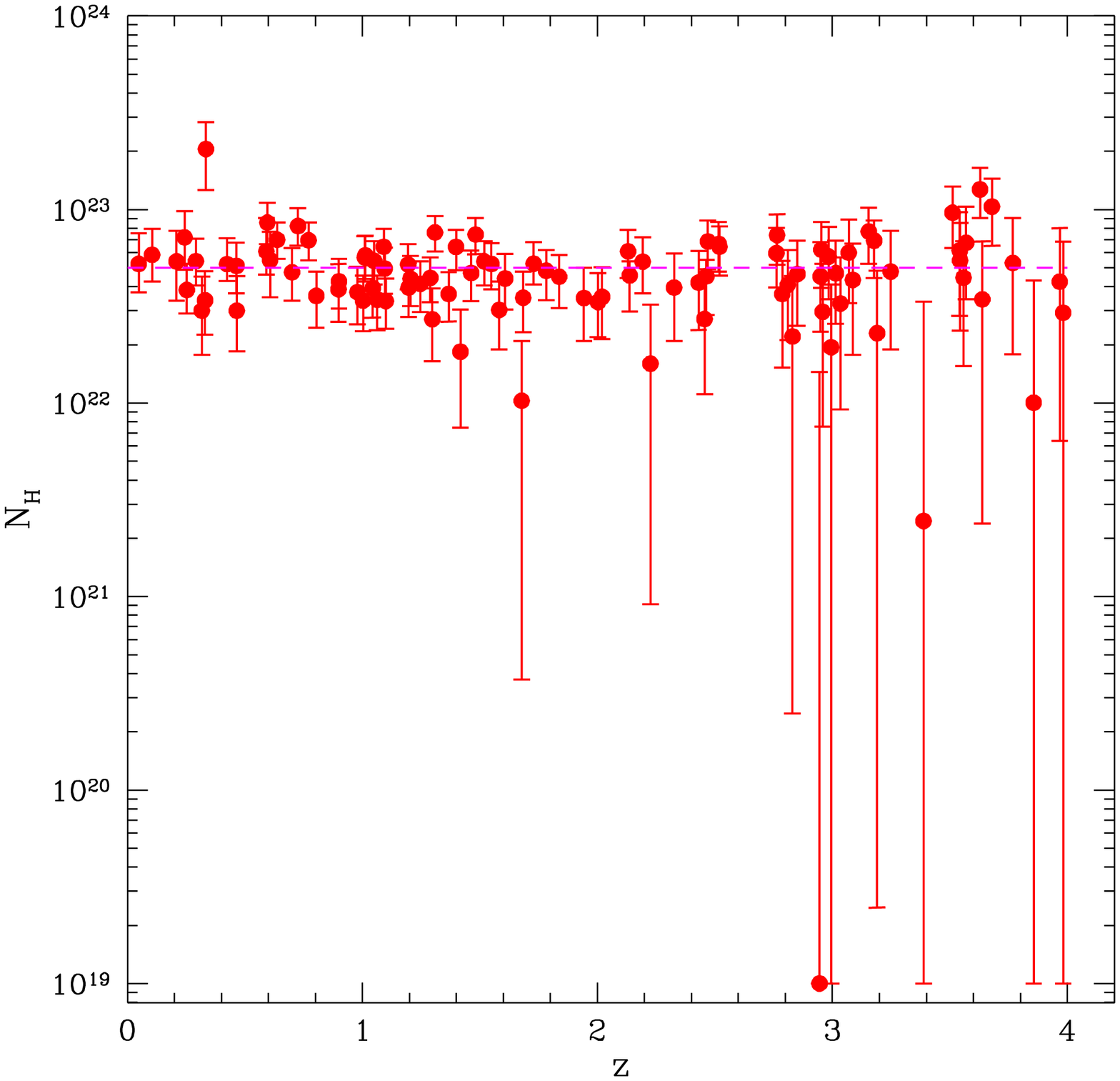}\includegraphics{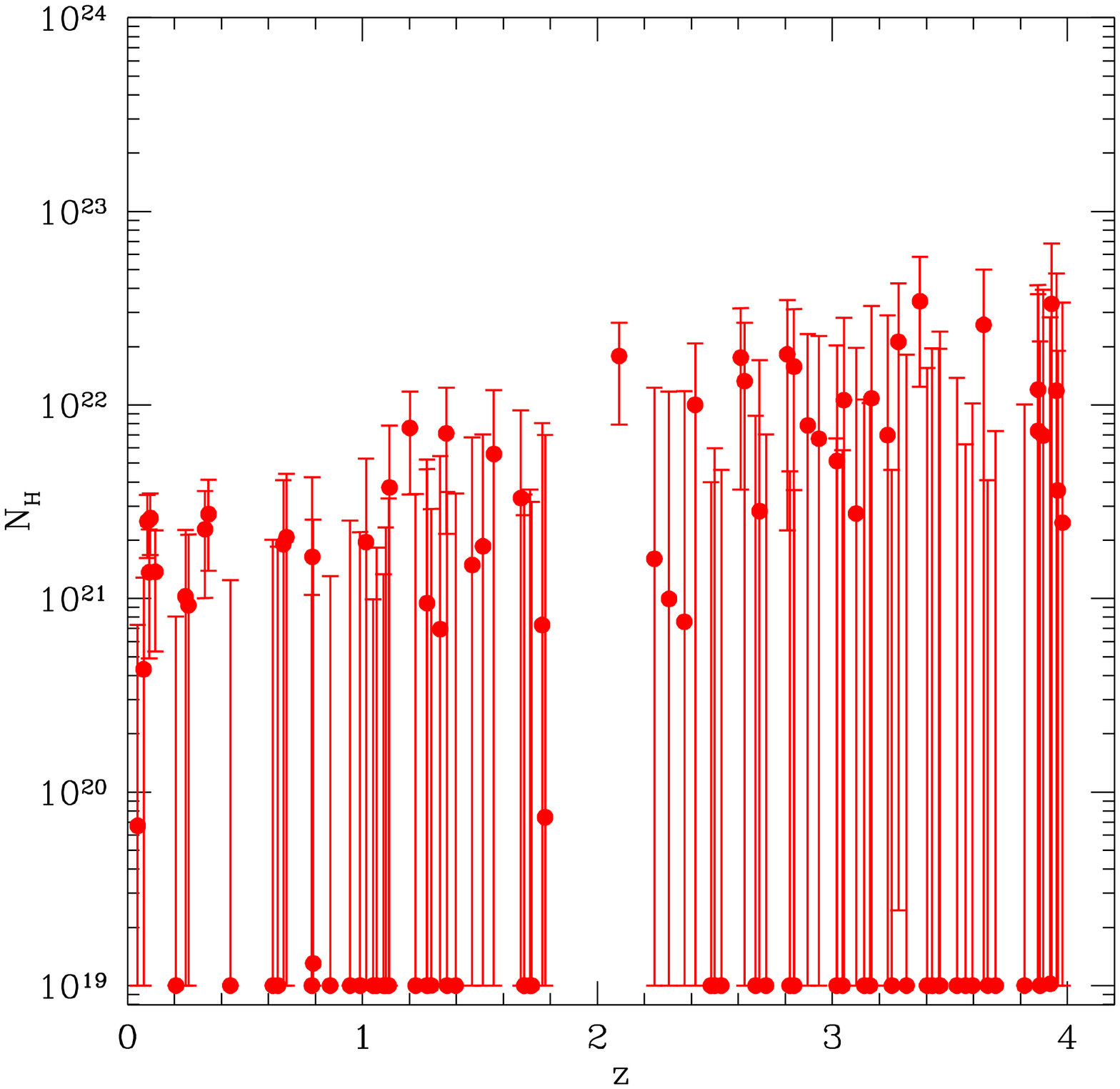}}
\caption{Best fit values of $N_H$ as a function of redshift for
simulated sources with input values $N_H = 5\times 10^{22}$ cm$^{-2}$
(left panel) and $N_H = 10^{21}$ cm$^{-2}$ (right panel).  Error bars
correspond to 1 $\sigma$.  The typical net detected counts are 280
(left panel) and 470 (right panel).  }
\label{nh_sim2}
\end{figure}

Therefore, the sample of BLAGN we discuss is consistent with no
absorption in average (since $N_H=0$ for all the BLAGN at 2 $\sigma$
c.l.).  However the number of sources with best fit value $N_H >
10^{22}$ cm$^{-2}$ among the BLAGN is 7, and it is larger than that
expected for negligible absorption on the basis of the simulations
(which is about 3).  We conclude that some of the high--z BLAGN do
have absorption at the level of $N_H \sim 10^{21}$ cm$^{-2}$, still
consistent with that found in Type I AGN.  Values typical of Type II
AGN, larger than $10^{22}$ cm$^{-2}$, would have been detected, while
values around $10^{21}$ cm$^{-2}$ are compatible with what we found in
the data (see Figure \ref{nh_sim2}).  This picture is confirmed by the
analysis of the XMM data for these sources (see discussion in \S 7).

To summarize, we find that the effect of an artificially increasing
measure of intrinsic absorption with redshift is under control in our
sample.  A complete removal of the spurious trend would need extensive
simulations and several iterations, since the effect depends on the
actual behaviour of $N_H$ with redshift.  Such an approach goes beyond
the scope of this Paper, but it should be adopted for larger AGN
samples.

\end{document}